\documentclass[a4paper,12pt]{article}

\usepackage{amsmath}
\usepackage[psamsfonts]{amssymb}
\usepackage[mathscr]{euscript}

\usepackage{bm}

\usepackage{cite}
\usepackage[dvips]{graphicx}

\makeatletter
\@addtoreset{equation}{section}
\makeatother

\begin{document} 

\begin{titlepage}

\hrule 
\leftline{}
\leftline{Chiba Univ. Preprint
          \hfill   \hbox{\bf CHIBA-EP-132}}
\leftline{\hfill   \hbox{hep-th/0111256}}
\leftline{\hfill   \hbox{November 2001}}
\vskip 5pt
\hrule 
\vskip 1.0cm
\centerline{\large\bf 
Renormalizing a BRST-invariant composite operator 
} 
\vskip 0.5cm
\centerline{\large\bf  
of mass dimension 2 in Yang-Mills theory
}
\vskip 0.5cm
\centerline{\large\bf  
}

\vskip 0.5cm

\centerline{{\bf 
K.-I. Kondo,$^{1{}^{\dagger}{}^{\ddagger}}$
 T. Murakami,$^{2{}^{\ddagger}}$ T. Shinohara$^{3{}^{\ddagger}}$
and T. Imai,$^{4{}^{\ddagger}}$
}}  
\vskip 1cm
\begin{description}
\item[]{\it \centerline{  
${}^{\dagger}$Department of Physics, Faculty of Science, 
Chiba University,  Chiba 263-8522, Japan}
  }
\item[]{\it 
${}^{\ddagger}$Graduate School of Science and Technology, 
Chiba University, Chiba 263-8522, Japan
  }
\end{description}

\begin{abstract}
 We discuss the renormalization of a BRST and anti-BRST invariant composite operator of mass dimension 2 in Yang-Mills theory with the general BRST and anti-BRST invariant gauge fixing term of the Lorentz type.
The interest of this study stems from a recent claim that the non-vanishing vacuum condensate of the composite operator in question can be an origin of mass gap and quark confinement in any manifestly covariant gauge, as proposed by one of the authors.  
First, we obtain the renormalization group flow of the Yang-Mills theory.
Next, we show the multiplicative renormalizability of the composite operator and that the BRST and anti-BRST invariance of the bare composite operator is preserved under the renormalization.
Third, we perform the operator product expansion of the gluon and ghost propagators and obtain the Wilson coefficient corresponding to the vacuum condensate of mass dimension 2.
Finally, we discuss the connection of this work with the previous works and argue the physical implications of the obtained results.

\end{abstract}


\vskip 0.2cm
\hrule  
\vskip 0.2cm
${}^1$ 
  E-mail:  {\tt kondo@cuphd.nd.chiba-u.ac.jp}

${}^2$ 
  E-mail:  {\tt tom@cuphd.nd.chiba-u.ac.jp}

${}^3$ 
  E-mail:  {\tt sinohara@cuphd.nd.chiba-u.ac.jp}

${}^4$ 
  E-mail:  {\tt takahito@physics.s.chiba-u.ac.jp}

\vskip 0.2cm  

\par 
\par\noindent
\vskip 0.5cm


\vskip 0.5cm

\newpage
\pagenumbering{roman}
\tableofcontents

\vskip 0.5cm  



\end{titlepage}


\pagenumbering{arabic}

\section{Introduction}

It is still a challenging and unsolved problem to prove quark confinement in the framework of quantum chromodynamics (QCD).
A very beginning question in deriving quark confinement is in what sense quark is confined?
A simple criterion of quark confinement which has been widely used so far is the area law decay of the Wilson loop (defined by the vacuum expectation value of the Wilson loop operator).  
The area law implies the presence of a linear piece $\sigma r$ proportional to the interquark distance $r$ in the static interquark potential $V(r)$.
The dual superconductivity of QCD vacuum \cite{dualsuper} is one of the most promising mechanisms of quark confinement in compatible with this picture.  
However, it is well known that this criterion is not so useful in the presence of dynamical matter, since the interquark force is screened by a pair of quark and anti-quark created from the vacuum and the linear piece no longer appears in the potential.
\par
In the previous paper \cite{Kondo01}, one of the authors (K.-I. K.) has proposed a non-vanishing vacuum condensate 
$\langle \mathcal{O} \rangle$
of mass dimension 2 as the origin of mass gap and quark confinement in Yang-Mills theory.  The proposed composite operator of mass dimension 2 is given by
\begin{equation}
 \mathcal{O} :={1 \over \Omega^{(D)}} \int d^D x \ \text{tr} \left[ 
 {1 \over 2} \mathscr{A}_\mu(x) \cdot \mathscr{A}_\mu(x) + \alpha i \bar{\mathscr{C}}(x) \cdot \mathscr{C}(x) 
\right] ,
\label{combi}
\end{equation}
where $\mathscr{A}_\mu$ is the gauge field, $\mathscr{C}$ ($\bar{\mathscr{C}}$) is the ghost (anti-ghost) field and $\Omega^{(D)}$ denotes the volume of the $D$-dimensional spacetime.
It has been shown \cite{Kondo01} that the composite operator $\mathcal{O}$ is invariant under the Becchi-Rouet-Stora-Tyutin (BRST) \cite{BRST} and anti-BRST \cite{antiBRST} transformations in the manifestly Lorentz covariant gauge, especially in the most general%
\footnote{The precise definition of `the most general' is stated later in the text.  Roughly speaking, the most general Lorentz gauge is obtained by imposing both the BRST and anti-BRST invariance for the gauge fixing term which corresponds to the Lorentz gauge 
$\partial^\mu \mathscr{A}_\mu(x)=0$.  The resulting gauge fixing term has two parameters.  The conventional Lorentz gauge is obtained as a special choice of the parameters.
}
 Lorentz gauge \cite{CF76,HL81,BT81,BT82,Baulieu85,DJ82} and the Maximal Abelian (MA) gauge \cite{tHooft81,KondoI,KondoII,KondoIV,KondoV,KondoVI,KS01,SIK01}.  
In (\ref{combi}), the trace is taken over the broken generators of the Lie algebra $\mathcal{G}$ of the original group $G$ when the original gauge group $G$ is broken to $H$ by a local gauge fixing condition chosen, i.e, $G$ itself in the Lorentz gauge and $G/H$ in the  MA gauge corresponding to the maximal torus group $H$ of $G$.  
Especially, in the limit $\alpha \rightarrow 0$ (which we call the Landau gauge), the composite operator reduces to 
$
 \mathcal{O} =(\Omega^{(D)})^{-1} \int d^D x \ \text{tr}  \left[
 {1 \over 2} \mathscr{A}_\mu(x) \cdot \mathscr{A}_\mu(x) \right]
$
and hence it becomes gauge invariant, since the contribution from the ghost and anti-ghost disappears.  
The vacuum condensate includes the ghost condensation proposed in the MA gauge \cite{Schaden99,KS00} and reduces to the gluon condensation proposed recently by several authors \cite{Boucaudetal00,Boucaudetal01,GSZ01,GZ01}, see also \cite{AZ98,CGPZ00}.
\par
The physical implication of the existence of such a condensate 
$\langle \mathcal{O} \rangle$
has been argued based on the operator product expansion (OPE) of the gluon and ghost propagators (2-point functions) and the vertex function (3-point function) \cite{Boucaudetal00,GZ01,Kondo01}.  However, the actual calculation has been performed within the tree level.  
\par
In order for such a proposal to be meaningful, it is very indispensable to show that the whole strategy to derive quark confinement based on the novel vacuum condensate survives the renormalization.  
In view of this, we focus on the renormalization of the composite operator (\ref{combi}).
The main purpose of this paper is to examine whether or not the composite operator in the integrand of $\mathcal{O}$ is renormalizable.  In addition, we must clarify the meaning of the BRST and anti-BRST symmetry in the renormalized theory.  We examine whether or not the renormalized composite operator $\mathcal{O}^{\rm R}$ is invariant under the renormalized BRST and anti-BRST transformation.  
If this is the case, the proposed composite operator of mass dimension 2 and the corresponding vacuum condensate can have a definite physical meaning. 
The analysis of this paper is restricted to the most general Lorentz gauge fixing, since the analysis of the MA gauge is more involved and hence the result is to be reported in a separate paper \cite{KIMS01}.

\par
In the most general Lorentz gauge, the multiplicative renormalizability of the Yang-Mills theory has been worked out by Baulieu and Thierry-Mieg \cite{BT82} by making use of the Slavnov-Taylor identities characterizing the BRST and anti-BRST invariance of the theory (See textbooks and reviews, e.g., \cite{Collins84,PT84,Kugo89,Zinn-Justin96,Weinberg96,Callan76,Lee76}
).  
In the course of renormalizing the composite operator, however, there is a subtle problem of the operator mixing.  In order to discuss the renormalization of a composite operator, we must take into account all the contributions coming from all the other composite operators of the same mass dimension and the same symmetry property.
In the OPE, the Wilson coefficient corresponding to arbitrary vacuum condensate can be calculated by the perturbation theory. 
In the usual Lorentz gauge, the Wilson coefficient associated with the ghost condensate $\langle \bar{\mathscr{C}} \cdot \mathscr{C}\rangle$ in the OPE of the propagator vanishes identically due to a special property of the 3-point gluon-ghost-anti--ghost vertex as pointed out in \cite{LS88}.
In the most general Lorentz gauge \cite{BT82,Baulieu85}, however, we show in this paper that the operator mixing between two composite operators, ${1 \over 2} \mathscr{A}_\mu \cdot \mathscr{A}_\mu$ and $i \bar{\mathscr{C}} \cdot \mathscr{C}$, of mass dimension 2 does exist in general due to the presence of four-ghost interaction (except for the case which is reduced to the conventional Lorentz gauge).  We explicitly calculate the matrix of renormalization factors of the composite operator in the one-loop level.  
\par
 For the Landau gauge, the vacuum condensate of mass dimension 2 in Yang-Mills theory is nothing but the gluon pair condensation.  A possibility of gluon pair condensation was already suggested from the existence of the tachyon pole in the two gluon channel by solving approximately the Bethe-Salpeter equation, see e.g. \cite{Fukuda78} and \cite{GM82}. 
A gluon pair can be identified as a Cooper pair, that is a bound state caused by the attractive force.   
Hence the gluon condensation is regarded as the Bose condensation of the gluon with spin 1.   
A remarkable point of our treatment different from the previous one is to retain the manifest Lorentz covariance and gauge (or BRST and anti-BRST) invariance.  
Hence the introduction of ghost field is indispensable in this approach.  It is important to clarify how the inclusion of the ghost influences the dynamics of gluon to recover the gauge invariance.   
This paper is a preliminary work toward the complete understanding of this problem.

\par
Another purpose of this paper is to point out that the composite operator discussed above
has the analog in the Abelian gauge theory, especially, quantum electrodynamics (QED).  
This suggests that a confinement phase can exist even in QED probably in the strong coupling region \cite{Miransky85,BLL86,KMY89,KSY91}.
In QED, the vacuum condensate in question is reduced in the Landau gauge to the photon pairing.
The photon pairing has also been suggested long ago by solving the Cooper equation, see \cite{Fukuda89,IF91}.
From quite a different viewpoint, one of the authors \cite{KondoIII} discussed the existence of a confinement phase in QED based on the total QED Lagrangian with the BRST and anti-BRST invariant gauge fixing term which is identical to the usual Lagrangian in the Lorentz gauge up to a total derivative term.  An advantage of rewriting the the gauge fixing part of the Lagrangian into the BRST and anti-BRST exact form is that the hidden supersymmetry becomes manifest and that the gauge-fixing part in four spacetime dimensions is reduced to the $O(2)$ non-linear sigma model in two spacetime dimensions owing to the Parisi-Sourlas dimensional reduction.%
\footnote{
This formulation has been applied to QED at finite temperature and a new confining phase is claimed to exist, see \cite{Yoshida01} and references therein.
}
  In view of this, the ghost is indispensable in this approach even for the Abelian gauge theory where the ghost decouples and is usually considered to be unnecessary.
In the analysis of quark confinement, it is most important to understand the origin of the scale or the mechanism of mass generation which was not so clear in the previous treatments.   
The detailed analysis of this issue will be reported in a subsequent paper.

\par
This paper is organized as follows.
In section 2, we summarize the BRST and anti-BRST transformations and their properties which are necessary in the following analyses.  

In section 3, we examine how the renormalization in QED is performed so as to preserve the BRST and anti-BRST symmetry.  
This section is a preliminary step for dealing with the non-Abelian gauge theory in the subsequent sections.  

In section 4, we consider the most general Lagrangian of the Yang-Mills theory which has manifest Lorentz covariance, global gauge invariance and BRST and anti-BRST symmetry.  The gauge fixing term contains two gauge fixing parameters.  We give Feynman rules of this theory and calculate the renormalization constants in the one-loop level.  Although some materials in this section are a reconfirmation of the results obtained by Baulieu and Thierry-Mieg \cite{BT82}, it is necessary to make this paper self-contained and to give basic ingredients in the subsequent sections. 

In section 5, we obtain the renormalization group flow in the parameter space of the theory.  To one-loop order, e specify the location of the fixed points and obtain the equation of the lines of connecting the fixed points.  

In section 6, we discuss the main subject of this paper; the renormalization of the composite operator $\mathcal{O}$ of mass dimension 2.  First, we show when the composite operator $\mathcal{O}$ is both BRST and anti-BRST invariant.
Next, we evaluate the renormalization of $\mathcal{O}$ by taking into account the mixing of the operators with the same mass dimension and the same symmetry.  To the best of our knowledge, the renormalization of the composite operator of mass dimension 2 has not been fully discussed except for a special case, i.e., the Landau gauge in the conventional Lorentz gauge fixing \cite{Boucaudetal01}.

In section 7, we perform the operator product expansion of the gluon and ghost propagators and obtain the Wilson coefficient associated with the vacuum condensates in question.

In the final section, we give the conclusion of this paper and discuss the future directions of our research.  
In Appendix, we give some of the calculations omitted in the text. 

\newpage
\section{BRST and anti-BRST transformation}

We consider the general non-Abelian gauge theory with a gauge group $G$.
In the following we use the notation:
\begin{equation}
 F \cdot G :=F^A G^A, \quad 
 F^2 :=F \cdot F , \quad 
 (F \times G)^A :=f^{ABC}F^B G^C ,
\end{equation}
where $f^{ABC}$ are the structure constants of the Lie algebra $\mathscr{G}$ of the gauge group $G$.
\par
For the non-Abelian gauge theory, we define the BRST transformation by
\begin{subequations}
\begin{align}
 \bm{\delta}_{\rm B} \mathscr{A}_\mu(x)
   & =\mathscr{D}_\mu[\mathscr{A}]\mathscr{C}(x)
    :=\partial_\mu \mathscr{C}(x)
      + g (\mathscr{A}_\mu(x) \times \mathscr{C}(x)) , \\
 \bm{\delta}_{\rm B} \mathscr{C}(x)
   & =-{1 \over 2}g(\mathscr{C}(x) \times \mathscr{C}(x)) , \\
 \bm{\delta}_{\rm B} \bar{\mathscr{C}}(x)
   & =i \mathscr{B}(x) , \\
 \bm{\delta}_{\rm B} \mathscr{B}(x)
   &=0 ,
\label{BRST1}
\end{align}
\end{subequations}
where 
$\mathscr{A}_\mu, \mathscr{B}, \mathscr{C}$ and $\bar{\mathscr{C}}$ are the non-Abelian gauge field, the Nakanishi-Lautrup (NL) auxiliary field, the Faddeev-Popov (FP) ghost and anti-ghost fields respectively.
Another BRST transformation, i.e., anti-BRST transformation~\cite{antiBRST} is  defined by
\begin{subequations}
\begin{align}
 \bar{\bm{\delta}}_{\rm B} \mathscr{A}_\mu(x) &=
\mathscr{D}_\mu[\mathscr{A}] \bar{\mathscr{C}}(x) :=
\partial_\mu \bar{\mathscr{C}}(x) + g (\mathscr{A}_\mu(x) \times \bar{\mathscr{C}}(x)) , \\
 \bar{\bm{\delta}}_{\rm B} \bar{\mathscr{C}}(x) &=-{1 \over 2}g(\bar{\mathscr{C}}(x) \times \bar{\mathscr{C}}(x)) ,
 \\
 \bar{\bm{\delta}}_{\rm B} \mathscr{C}(x) &=i \bar{\mathscr{B}}(x) ,
\\
 \bar{\bm{\delta}}_{\rm B} \bar{\mathscr{B}}(x) &=0 ,
\label{BRST2}
\end{align}
\end{subequations}
where%
\footnote{
The last transformation is equivalent to
\begin{equation}
  \bar{\bm{\delta}}_{\rm B} \mathscr{B}(x) =-g \bar{\mathscr{C}}(x) \times \mathscr{B}(x) .
\end{equation}
}
$\bar{\mathscr{B}}$ is defined by
\begin{equation}
  \bar{\mathscr{B}}(x) =-\mathscr{B}(x) + ig (\mathscr{C}(x) \times \bar{\mathscr{C}}(x)) .
\end{equation}
The BRST and anti-BRST transformations are nilpotent and they anti-commute:
\begin{equation}
  \bm{\delta}_{\rm B} \bm{\delta}_{\rm B} \equiv 0 , \quad
\bar{\bm{\delta}}_{\rm B} \bar{\bm{\delta}}_{\rm B} \equiv 0 , \quad
 \bm{\delta}_{\rm B} \bar{\bm{\delta}}_{\rm B} + \bar{\bm{\delta}}_{\rm B} \bm{\delta}_{\rm B} \equiv 0 .
\end{equation}
\par
For the Abelian gauge theory, 
the BRST transformation reads
\begin{subequations}
\begin{align}
 \bm{\delta}_{\rm B} a_\mu(x) &=
 \partial_\mu C(x),  \\
 \bm{\delta}_{\rm B} C(x) &=0 ,
\\
 \bm{\delta}_{\rm B} \bar{C}(x) &=i B(x) ,
\\
 \bm{\delta}_{\rm B} B(x) &=0 ,
\label{BRSTA1}
\end{align}
\end{subequations}
where 
$A_\mu, B, C$ and $\bar{C}$ are the Abelian gauge field, the NL auxiliary field, the FP ghost and anti-ghost fields respectively.
The anti-BRST transformation is reduced to
\begin{subequations}
\begin{align}
 \bar{\bm{\delta}}_{\rm B} a_\mu(x) &=
\partial_\mu C(x)  , \\
 \bar{\bm{\delta}}_{\rm B} \bar{C}(x) &=0 ,
 \\
 \bar{\bm{\delta}}_{\rm B} C(x) &=i \bar{B}(x) ,
\\
 \bar{\bm{\delta}}_{\rm B} \bar{B}(x) &=0 ,
\label{BRSTA2}
\end{align}
\end{subequations}
where $\bar{B}$ is defined by
\begin{equation}
  \bar{B}(x) =-B(x)  .
\end{equation}

\section{QED in the Lorentz gauge}
As a warming-up problem, we consider the  quantum electrodynamics (QED).  
As is well known, the total Lagrangian of  QED is given by 
\begin{align}
  \mathscr{L}_{\rm QED}^{\rm tot} =- {1 \over 4} f^{\mu\nu} f_{\mu\nu} 
+ \bar{\psi} (i\gamma^\mu \partial_\mu - m)\psi
- e \bar{\psi} \gamma^\mu \psi a_\mu 
+ \mathscr{L}_{\rm GF+FP} , 
\end{align}
with a gauge-fixing (GF) plus FP ghost term $\mathscr{L}_{\rm GF+FP}$.
The explicit form of the GF+FP term depends on the gauge chosen.  In this paper we adopt the most familiar covariant gauge, i.e., the Lorentz gauge
\begin{equation}
  \partial^\mu a_\mu =0 .  
\end{equation}
Therefore, the GF+FP term is given by
\begin{equation}
  \mathscr{L}_{\rm GF+FP} =-i \bm{\delta}_{\rm B}  \left(  \bar{C} \partial^\mu  a_\mu + {\alpha \over 2}   \bar{C} B \right) 
=B \partial^\mu  a_\mu + {\alpha \over 2}   B^2 + i \bar{C} \partial^\mu \partial_\mu C .
\label{GFA0}
\end{equation}
Although the ghost and anti-ghost fields are free and decouple from other fields,  we have included them to study the relationship with the non-Abelian case which will be discussed in the next section.
\par
 As pointed out in \cite{KondoIII}, the GF+FP term (\ref{GFA0}) is rewritten into the BRST and anti-BRST exact form,
\begin{equation}
  \mathscr{L}_{\rm GF+FP} =i \bm{\delta}_{\rm B} \bar{\bm{\delta}}_{\rm B} \left( {1 \over 2} a_\mu a^\mu + {\alpha \over 2}i  \bar{C} C \right) .
\label{GFA1}
\end{equation}
In fact, this is cast into the form,
\begin{align}
  \mathscr{L}_{\rm GF+FP} &=i \bm{\delta}_{\rm B}  \left(   (\bar{\bm{\delta}}_{\rm B} a^\mu) a_\mu  - {\alpha \over 2}i  \bar{C} \bar{\bm{\delta}}_{\rm B} C \right) 
\nonumber\\
&=i \bm{\delta}_{\rm B}  \left(   \partial^\mu \bar{C} a_\mu - {\alpha \over 2}   \bar{C} B \right) ,
\end{align}
which agrees with (\ref{GFA0}) up to a total-derivative term.
\par
If the NL field $B$ is eliminated by performing the functional integration or by making use of the equation of motion, then we obtain
\begin{equation}
  \mathscr{L}_{\rm GF+FP}' =- {1 \over 2\alpha} (\partial^\mu  a_\mu)^2  + i \bar{C} \partial^\mu \partial_\mu C .
\label{GFA2}
\end{equation}
The on-shell BRST transformation is given by
\begin{subequations}
\begin{align}
 \bm{\delta}_{\rm B} a_\mu(x) &=
 \partial_\mu C(x),  \\
 \bm{\delta}_{\rm B} C(x) &=0 ,
\\
 \bm{\delta}_{\rm B} \bar{C}(x) &=-{i \over \alpha} \partial^\mu  a_\mu (x) ,
\label{BRSTA3}
\end{align}
\end{subequations}
while the on-shell anti-BRST transformation is 
\begin{subequations}
\begin{align}
 \bar{\bm{\delta}}_{\rm B} a_\mu(x) &=
\partial_\mu \bar{C}(x)  , \\
 \bar{\bm{\delta}}_{\rm B} \bar{C}(x) &=0 ,
 \\
 \bar{\bm{\delta}}_{\rm B} C(x) &=+ {i \over \alpha} \partial^\mu  a_\mu (x) .
\label{BRSTA4}
\end{align}
\end{subequations}
The GF+FP Lagrangian $\mathscr{L}_{\rm GF+FP}'$ and the total Lagrangian $\mathscr{L}_{\rm QED}^{\rm tot}$ with $\mathscr{L}_{\rm GF+FP}'$ are separately invariant under the on-shell BRST and on-shell anti-BRST transformations.
The nilpotency of the on-shell BRST and anti-BRST transformation is realized only when the equation of motion for the ghost and anti-ghost field is used, since 
\begin{subequations}
\begin{align}
 (\bm{\delta}_{\rm B})^2 a_\mu(x) &=0,  \\
 (\bm{\delta}_{\rm B})^2 C(x) &=0 ,
\\
 (\bm{\delta}_{\rm B})^2 \bar{C}(x) &=-{i \over \alpha} \partial^\mu  \partial_\mu C(x) ,
\label{BRSTA5}
\end{align}
\end{subequations}
and
\begin{subequations}
\begin{align}
 (\bar{\bm{\delta}}_{\rm B})^2 a_\mu(x) &=0 , 
 \\
 (\bar{\bm{\delta}}_{\rm B})^2 C(x) &=+ {i \over \alpha} \partial^\mu  \partial_\mu \bar{C}(x) ,
\\
 (\bar{\bm{\delta}}_{\rm B})^2 \bar{C}(x) &=0 .
\label{BRSTA6}
\end{align}
\end{subequations}
Moreover, we obtain the similar result for the anti-commutability:
\begin{subequations}
\begin{align}
 (\bm{\delta}_{\rm B} \bar{\bm{\delta}}_{\rm B} + \bar{\bm{\delta}}_{\rm B} \bm{\delta}_{\rm B} ) a_\mu(x) &=0 , 
 \\
 (\bm{\delta}_{\rm B} \bar{\bm{\delta}}_{\rm B} + \bar{\bm{\delta}}_{\rm B} \bm{\delta}_{\rm B} ) C(x) &=- {i \over \alpha} \partial^\mu  \partial_\mu \bar{C}(x) ,
 \\
 (\bm{\delta}_{\rm B} \bar{\bm{\delta}}_{\rm B} + \bar{\bm{\delta}}_{\rm B} \bm{\delta}_{\rm B} ) \bar{C}(x) &=+ {i \over \alpha} \partial^\mu  \partial_\mu \bar{C}(x) .
\label{BRSTA7}
\end{align}
\end{subequations}

\par
Now we define the composite operator $\mathcal{O}$ of mass dimension 2 by
\begin{equation}
  \mathcal{O} :={1 \over \Omega^{(D)}} \int d^Dx \mathcal{Q}(x), \quad
  \mathcal{Q}(x) :={1 \over 2} a_\mu(x) a^\mu(x) +  \alpha i  \bar{C}(x) C(x) .
\end{equation}
This composite operator is BRST and anti-BRST invariant, since
\begin{equation}
  \bm{\delta}_{\rm B} \mathcal{Q}(x)  =\partial^\mu (a_\mu(x) C(x)) ,
\quad 
  \bar{\bm{\delta}}_{\rm B} \mathcal{Q}(x)  =\partial^\mu (a_\mu(x) \bar{C}(x)) .
\end{equation}
\par
We consider the renormalization of the composite operator $\mathcal{Q}$.
The Abelian case is very simple due to the trivial renormalization factors $Z_{a^2}$, $Z_{CC}$ for the composite field ${1 \over 2}a^\mu a_\mu$ and $i\bar{C}C$.  Therefore, we have only to take into account the renormalization factor of the fundamental field, $a_\mu, C, \bar{C}$ and the gauge fixing parameter $\alpha$.  
QED is known to be multiplicatively renormalizable in the sense that the divergences are absorbed by introducing the renormalization factors in the following way.
\begin{align}
  \psi &=Z_2^{1/2} \psi^{\rm R} ,
\\
  a_\mu &=Z_3^{1/2} a_\mu^{\rm R} ,
\\
  C &=Z_C C^{\rm R}, \quad \bar{C} =Z_{\bar{C}} \bar{C}^{\rm R} ,
\\
  (B &=Z_3^{-1/2} B^{\rm R}),
\\
   m &=Z_m Z_2^{-1} m^{\rm R} ,
\\
  \alpha &=Z_\alpha \alpha^{\rm R} ,
\\
  e &=Z_1 Z_2^{-1} Z_3^{-1/2} e^{\rm R} .
\end{align}
The renormalization factors are not independent to each other.
In fact, the coupling constant is renormalized as
\begin{equation}
  e =Z_3^{-1/2} e^{\rm R}  ,
\end{equation}
as a consequence of the Ward relation:
\begin{equation}
   Z_1 =Z_2 .
\end{equation}
Moreover, the Ward-Takahashi identity yields
\begin{equation}
  Z_\alpha =Z_3 .
\end{equation}
The result of perturbative renormalization in QED is well known and can be seen in the text books.
A result:
\begin{equation}
 Z_C =Z_{\bar{C}} =1 ,
\end{equation}
means that both  ghost and anti-ghost are free and receive no renormalization in the perturbation theory (This is not the case in the non-Abelian case). Consequently, we arrive at the result that the composite operator is renormalized as
\begin{equation}
  \mathcal{Q} =Z_3 \mathcal{Q}^{\rm R}, \quad
  \mathcal{Q}^{\rm R} :={1 \over 2} a_\mu^{\rm R}(x) a^\mu{}^{\rm R}(x) +  \alpha^{\rm R} i  \bar{C}^{\rm R}(x) C^{\rm R}(x) .
\end{equation}
Therefore, the BRST invariant combination of two composite operators with mass dimension 2 is preserved under the renormalization.  
\par
In view of the above results, the renormalized BRST transformation is defined by
\begin{equation}
  \bm{\delta}_{\rm B}^{\rm R} =Z_3^{1/2} \bm{\delta}_{\rm B},
\quad
 \bar{\bm{\delta}}_{\rm B}^{\rm R} =Z_3^{1/2} \bar{\bm{\delta}}_{\rm B} .
\label{rBRS}
\end{equation}
This is shown as follows.
The Noether current of the BRST symmetry is obtained as
\begin{equation}
  J_{\rm B}^\mu =B \partial^\mu C - \partial^\mu B C 
- \partial_\nu (f^{\mu\nu} C) .
\end{equation}
The Noether charge, i.e. the BRST charge $Q_B$  as the generator of the BRST transformation 
\begin{equation}
  [ i \lambda Q_B, \Phi(x) ] =\lambda \bm{\delta}_{\rm B} \Phi(x) ,
  \quad
\end{equation}
is given by
\begin{equation}
  Q_{\rm B} =\int d^3x J_{\rm B}^0
            =\int d^3x [ B \partial^0 C - \partial^0 B C ] .
\end{equation}
In the similar way, the anti-BRST charge $\bar{Q}_{\rm B}$  can also be defined as the Noether charge for the anti-BRST transformation.
Therefore we can define the renormalized BRST charge $Q_{\rm B}^{\rm R}$ as
\begin{equation}
  Q_{\rm B}^{\rm R} =Z_3^{1/2} Q_{\rm B} =\int d^3x [ B^{\rm R} \partial^0 C^{\rm R} - \partial^0 B^{\rm R} C^{\rm R} ] .
\end{equation}
This ensures the renormalization of the BRST transformation (\ref{rBRS}).  The renormalized BRST transformation for the renormalized field has the same form as the bare BRST transformation for the bare field.
Thus, the composite operator $\mathcal{Q}$ is a BRST invariant and multiplicatively renormalizable operator for arbitrary gauge parameter $\alpha$.
The renormalized GF+FP term has the same form as the bare one:
\begin{equation}
  \mathscr{L}_{\rm GF+FP} =i \bm{\delta}_{\rm B}^{\rm R} \bar{\bm{\delta}}_{\rm B}^{\rm R} \left( {1 \over 2} a_\mu^{\rm R} a^\mu{}^{\rm R} + {\alpha^{\rm R} \over 2}i  \bar{C}^{\rm R} C^{\rm R} \right) .
\label{GFAf}
\end{equation}

\section{Yang-Mills theory in the most general Lorentz gauge}

\subsection{Lagrangian}

We consider the most general quantum Lagrangian density that is a local function of the fields, $\mathscr{A}_\mu^A$, $\mathscr{B}^A$, $\mathscr{C}^A$, $\bar{\mathscr{C}}^A$ and satisfies the conditions:
The Lagrangian is (1) of mass dimension 4, 
(2) Lorentz invariant,
(3a) BRST invariant,
(3b) anti-BRST invariant,
(4) Hermitian,
(5) of zero ghost number, 
(6) global gauge invariant, 
and the theory with this Lagrangian is 
(7) (multiplicative) renormalizable.
Here it is implicitly assumed that the Lagrangian is written as the polynomial of the fields, and that there are no higher derivative terms, since there is no intrinsic mass scale in the Yang-Mills theory.  It should be remarked that we have imposed BRST and anti-BRST invariance instead of gauge invariance (we do not require gauge invariance for the Lagrangian).  
Such a Lagrangian was given by Baulieu and Thierry-Mieg \cite{BT82,Baulieu85} as 
\begin{align}
\mathscr{L}_{\rm YM}^{\rm tot}
  =\,& -{1 \over 4} \alpha_1 \mathscr{F}_{\mu\nu}\cdot \mathscr{F}^{\mu\nu}  
     + \alpha_2 \epsilon_{\mu\nu\rho\sigma}  \mathscr{F}^{\mu\nu} \cdot \mathscr{F}^{\rho\sigma}  
\nonumber\\
   & + i \bm{\delta}_{\rm B} \bar{\bm{\delta}}_{\rm B} \left( \alpha_3 \mathscr{A}_\mu \cdot \mathscr{A}^\mu  + \alpha_4 \mathscr{C} \cdot \bar{\mathscr{C}} \right) 
+ {\alpha' \over 2} \mathscr{B} \cdot \mathscr{B}  ,
\end{align}
where $\alpha_i(i=1,2,3,4)$ is an arbitrary constant, and $\bm{\delta}_{\rm B}$ and $\bar{\bm{\delta}}_{\rm B}$ are the BRST and anti-BRST transformations.
The first term is the Yang-Mills Lagrangian, the second term is the topological term which is not discussed in this paper and omitted hereafter.  The first and the second terms are gauge invariant.  On the other hand, the third and the fourth terms are identified with the GF and FP term, since they break the gauge invariance of the Lagrangian.  After the rescaling of the parameters and the field redefinitions, we can cast the total Lagrangian of the Yang-Mills theory into the form,
\begin{equation}
  \mathscr{L}_{\rm YM}^{\rm tot} =- {1 \over 4} \mathscr{F}_{\mu\nu} \cdot \mathscr{F}^{\mu\nu}  + \mathscr{L}_{\rm GF+FP} ,
  \label{totalL}
\end{equation}
with the GF+FP term \cite{BT82,Baulieu85,DJ82}:
\begin{align}
  \mathscr{L}_{\rm GF+FP} &=i \bm{\delta}_{\rm B} \bar{\bm{\delta}}_{\rm B} \left( {1 \over 2} \mathscr{A}_\mu \cdot \mathscr{A}^\mu  -{\alpha \over 2}i \mathscr{C} \cdot \bar{\mathscr{C}} \right) 
+ {\alpha' \over 2} \mathscr{B} \cdot \mathscr{B} 
\\
&=-i \bm{\delta}_{\rm B} \left( - \partial_\mu \bar{\mathscr{C}} \cdot \mathscr{A}^\mu + {\alpha \over 2} \bar{\mathscr{C}} \cdot \mathscr{B} - {i \over 4} \alpha g \bar{\mathscr{C}} \cdot (\bar{\mathscr{C}} \times \mathscr{C})   \right) 
+ {\alpha' \over 2} \mathscr{B} \cdot \mathscr{B}  .
\label{GFglobal}
\end{align}
The final term is allowed for the renormalizability of the total Lagrangian and is written in either BRST exact or anti-BRST exact form,
\begin{equation}
  \mathscr{B} \cdot \mathscr{B} =-i\bm{\delta}_{\rm B}(\bar{\mathscr{C}} \cdot \mathscr{B})
=i\bar{\bm{\delta}}_{\rm B}(\mathscr{C} \cdot \mathscr{B}) .  
\end{equation}
However, the GF+FP term (\ref{GFglobal}) is simultaneously BRST and anti-BRST exact, i.e., $\bm{\delta}_{\rm B} \bar{\bm{\delta}}_{\rm B}(*)$, only if $\alpha'=0$.
If we impose one more condition, e.g., the FP ghost conjugation invariance,
\begin{equation}
 \mathscr{C}^A \rightarrow \pm \bar{\mathscr{C}}^A, \quad
 \bar{\mathscr{C}}^A \rightarrow \mp \mathscr{C}^A, \quad
 \mathscr{B}^A \rightarrow - \bar{\mathscr{B}}^A, \quad
 \bar{\mathscr{B}}^A \rightarrow - \mathscr{B}^A \quad
 (\mathscr{A}_\mu^A \rightarrow \mathscr{A}_\mu^A) ,
\end{equation}
the second term of (\ref{GFglobal}) is excluded, namely, only the choice $\alpha'=0$ is allowed.
\par
By performing the BRST and anti-BRST transformations, we obtain
\begin{align}
 \mathscr{L}_{\rm GF+FP} =\,& 
  {\alpha+\alpha' \over 2} \mathscr{B} \cdot \mathscr{B} - {\alpha \over 2} ig (\mathscr{C} \times \bar{\mathscr{C}}) \cdot \mathscr{B} 
+ \mathscr{B} \cdot \partial_\mu \mathscr{A}^\mu
\nonumber\\&
 + i \bar{\mathscr{C}} \cdot \partial_\mu \mathscr{D}^\mu[\mathscr{A}]\mathscr{C}
+ {\alpha \over 8}g^2 (\bar{\mathscr{C}} \times \bar{\mathscr{C}}) \cdot (\mathscr{C} \times \mathscr{C}) 
\\
=\,& {\alpha+\alpha' \over 2}\mathscr{B} \cdot \mathscr{B}
   - {\alpha \over 2} ig (\mathscr{C} \times \bar{\mathscr{C}}) \cdot \mathscr{B} 
+ \mathscr{B} \cdot \partial_\mu \mathscr{A}^\mu 
\nonumber\\&
+ i \bar{\mathscr{C}} \cdot \partial_\mu \mathscr{D}^\mu[\mathscr{A}]\mathscr{C}
+ {\alpha \over 4}g^2 (i \mathscr{C} \times \bar{\mathscr{C}}) \cdot (i \mathscr{C} \times \bar{\mathscr{C}}) .
\end{align}
The GF+FP term includes the ghost self-interaction where the strength 
is proportional to the parameter $\alpha$.
\par
When $\alpha=0$, this theory reduces to the usual Yang-Mills theory in the Lorentz type gauge fixing with the gauge fixing parameter $\alpha'$:
\begin{align}
 \mathscr{L}_{\rm GF+FP} &=
  {\alpha' \over 2} \mathscr{B} \cdot \mathscr{B} 
+ \mathscr{B} \cdot \partial_\mu \mathscr{A}^\mu + i \bar{\mathscr{C}} \cdot \partial_\mu \mathscr{D}^\mu[\mathscr{A}]\mathscr{C} .
\end{align}
This is consistent with the FP prescription.
\par
When $\alpha \not=0$, there exists a quartic ghost interaction which can not be implemented by the usual FP prescription.  Therefore we must go beyond the FP prescription.  The GF+FP term is further rewritten as
\begin{align}
 \mathscr{L}_{\rm GF+FP} 
=\,& 
- {1 \over 2\lambda}(\partial^\mu \mathscr{A}_\mu)^2
+ (1-\xi)i \bar{\mathscr{C}} \cdot \partial_\mu \mathscr{D}^\mu[\mathscr{A}]\mathscr{C}
+ \xi i \bar{\mathscr{C}} \cdot \mathscr{D}^\mu[\mathscr{A}] \partial_\mu \mathscr{C}
\nonumber\\&
+ {1 \over 2}\lambda \xi (1-\xi) g^2 (i \mathscr{C} \times \bar{\mathscr{C}}) \cdot (i \mathscr{C} \times \bar{\mathscr{C}}) 
\nonumber\\&
 + {\lambda \over 2} \left( \mathscr{B} + \lambda^{-1} \partial^\mu \mathscr{A}_\mu - \xi ig (\mathscr{C} \times \bar{\mathscr{C}}) \right)^2 
\\
=\,& - {1 \over 2\lambda}(\partial^\mu \mathscr{A}_\mu)^2
+ i \bar{\mathscr{C}} \cdot \partial_\mu \partial^\mu \mathscr{C}
- (1-\xi)g i  \mathscr{A}^\mu \cdot (\partial_\mu \bar{\mathscr{C}} \times \mathscr{C})
\nonumber\\&
+ \xi g i \mathscr{A}^\mu \cdot (\bar{\mathscr{C}} \times \partial_\mu \mathscr{C})
+ {1 \over 2}\lambda \xi (1-\xi) g^2 (i \mathscr{C} \times \bar{\mathscr{C}}) \cdot (i \mathscr{C} \times \bar{\mathscr{C}})
\nonumber\\&
 + {\lambda \over 2} \left( \mathscr{B} + \lambda^{-1} \partial^\mu \mathscr{A}_\mu - \xi ig (\mathscr{C} \times \bar{\mathscr{C}}) \right)^2  ,
 \label{GF+FP}
\end{align}
where we have defined the two parameters%
\footnote{
The parameters $\alpha, \alpha', \lambda, \xi$ in this paper corresponds respectively to the $\lambda_c, \lambda_b, \lambda, \alpha$ in \cite{Baulieu85} and $a, a', \lambda, \alpha/2$ in \cite{BT82}.
}
\begin{equation}
  \lambda :=\alpha+\alpha' , \quad
  \xi :={\alpha/2 \over \alpha+\alpha'} ={\alpha \over 2\lambda}.
\end{equation}
In this form, it is easy to eliminate the Nakanishi-Lautrup field $\mathscr{B}$. 
We call the gauge (\ref{GF+FP}) the most general Lorentz gauge hereafter.

\subsection{Feynman rules} %
We obtain the following Feynman rules for the Yang-Mills theory of the Lagrangian (\ref{totalL}) with (\ref{GF+FP}) where the NL field is eliminated.
\subsubsection{Propagators} %
\unitlength=0.001in
\begin{enumerate}

\item[(a)] gluon propagator:
\begin{equation}
\begin{array}{c}
\begin{picture}(1200,400)(-300,-100)
   \put(0,0){\includegraphics[height=.5cm]{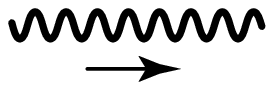}}%
   \put(-330,110){\mbox{$A,\mu$}}%
   \put(680,110){\mbox{$B,\nu$}}%
   \put(230,-50){\mbox{$p$}}%
\end{picture}
\end{array}
  =iD_{\mu\nu}^{AB}
  =-\frac i{p^2}
    \left[g_{\mu\nu}-(1-\lambda)\frac{p_\mu p_\nu}{p^2}\right]\delta^{AB} .
\label{eq:propagator of off-diagonal gluon}
\end{equation}

\item[(b)] ghost propagator:
\begin{equation}
\begin{array}{c}
\begin{picture}(900,250)(-200,-100)
   \put(-50,0){\includegraphics[height=.2cm]{ghost.eps}}%
   \put(-200,0){\mbox{$A$}}%
   \put(630,0){\mbox{$B$}}%
   \put(200,-100){\mbox{$p$}}%
\end{picture}
\end{array}
=i G^{AB}
  =-\frac1{p^2}\delta^{AB}.
\end{equation}
\end{enumerate}

\subsubsection{Three-point vertices} %
\begin{enumerate}
\item[(c)] Three-gluon vertex:
\begin{equation}
\begin{array}{c}
\begin{picture}(1100,800)
   \put(0,0){\includegraphics[height=2cm]{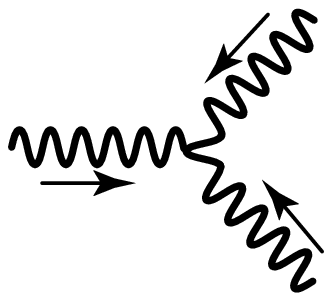}}%
   \put(150,200){\mbox{$p$}}%
   \put(580,750){\mbox{$q$}}%
   \put(880,150){\mbox{$r$}}%
   \put(100,500){\mbox{$A,\mu$}}%
   \put(860,600){\mbox{$B,\rho$}}%
   \put(420,20){\mbox{$C,\sigma$}}%
\end{picture}
\end{array}
\textstyle
 =gf^{ABC}
  \left[(q-r)_\mu g_{\rho\sigma}
        +(r-p)_\rho g_{\sigma\mu}
        +(p-q)_\sigma g_{\mu\rho}
  \right].
\end{equation}

\item[(d)] Gluon--ghost--anti-ghost vertex:
\begin{equation}
\begin{array}{c}
\begin{picture}(800,800)
   \put(0,0){\includegraphics[height=2cm]{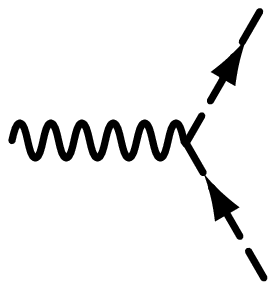}}%
   \put(570,700){\mbox{$p$}}%
   \put(700,150){\mbox{$q$}}%
   \put(100,500){\mbox{$C,\mu$}}%
   \put(700,600){\mbox{$A$}}%
   \put(500,30){\mbox{$B$}}%
\end{picture}
\end{array}
  =igf^{ABC}\left[\textstyle\xi(p-q)-p\right]^\mu.
\label{ggagv}
\end{equation}

\end{enumerate}

\subsubsection{Four-point vertices} %
\begin{enumerate}
\item[(e)] Four-gluon vertex:
\begin{align}
\begin{array}{c}
\begin{picture}(1200,800)(-300,0)%
   \put(0,0){\includegraphics[height=2cm]{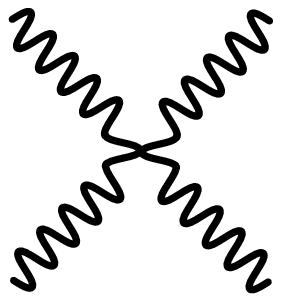}}%
   \put(-280,600){\mbox{$A,\mu$}}%
   \put(-280,100){\mbox{$B,\nu$}}%
   \put(730,600){\mbox{$C,\rho$}}%
   \put(730,100){\mbox{$D,\sigma$}}%
\end{picture}
\end{array}
\textstyle
    =-i2g^2\bigl(\,&f^{EAB}f^{ECD}I_{\mu\nu,\rho\sigma}
                    \nonumber\\[-1.2\baselineskip]
                   &+f^{EAC}f^{EBD}I_{\mu\rho,\nu\sigma}
                    \nonumber\\
                   &+f^{EAD}f^{EBC}I_{\mu\sigma,\nu\rho}\,\bigr),
\end{align}
where
$I_{\mu\nu,\rho\sigma}
  :=(g_{\mu\rho}g_{\nu\sigma}
     -g_{\mu\sigma}g_{\nu\rho})/2
$
.

\item[(f)] Four-ghost vertex:
\begin{equation}
\begin{array}{c}
\begin{picture}(900,800)(-100,0)%
   \put(0,0){\includegraphics[height=2cm]{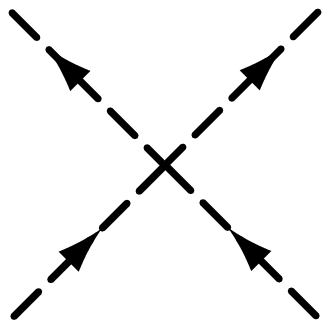}}%
   \put(-50,580){\mbox{$A$}}%
   \put(-50,130){\mbox{$C$}}%
   \put(700,600){\mbox{$B$}}%
   \put(700,100){\mbox{$D$}}%
\end{picture}
\end{array}
  = -i\lambda\xi(1-\xi)
     g^2\left(f^{ACE}f^{BDE}
              -f^{ADE}f^{BCE}\right).
\end{equation}

\end{enumerate}

\subsection{Multiplicative renormalization}

It has been proved by Baulieu and Thierry-Mieg \cite{BT82} based on the mathematical induction that the Yang-Mills theory in the most general Lorentz gauge (\ref{GF+FP}) is multiplicatively renormalizable.  
We introduce the renormalization constant (or renormalization factor) for the field:
\begin{align}
  \mathscr{A}_\mu =Z_A^{1/2} \mathscr{A}_\mu^{\rm R}, \quad
  \mathscr{C}  =Z_C^{1/2} \mathscr{C}^{\rm R} , \quad
  \bar{\mathscr{C}} =Z_C^{1/2} \bar{\mathscr{C}}^{\rm R} , \quad
  \mathscr{B} =Z_B^{1/2} \mathscr{B}^{\rm R} 
=Z_C Z_A^{-1/2} \mathscr{B}^{\rm R} ,
\label{rc1}
\end{align}
and for the parameters:
\begin{align}
  \lambda =Z_\lambda \lambda_{\rm R} , \quad
  \xi =Z_\xi \xi_{\rm R} , \quad
  g =Z_g g_{\rm R} .
\label{rc2}
\end{align}

By substituting (\ref{rc1}) and (\ref{rc2}) into the bare Lagrangian, we obtain the total Lagrangian written in terms of the renormalized fields, renormalized parameters and the renormalization factors:
\begin{align}
 \mathscr{L}_{\rm YM}^{\rm tot} 
=\,& 
- {1 \over 4} Z_A (\partial_\mu \mathscr{A}_\nu^{\rm R} - \partial_\nu \mathscr{A}_\mu^{\rm R} + Z_g Z_A^{1/2} g_{\rm R} \mathscr{A}_\mu^{\rm R} \times \mathscr{A}_\nu^{\rm R} )^2
\nonumber\\&
 - {1 \over 2\lambda_{\rm R}} Z_A Z_\lambda^{-1}(\partial^\mu \mathscr{A}_\mu^{\rm R})^2
+ i Z_C \bar{\mathscr{C}}^{\rm R} \cdot \partial_\mu \partial^\mu \mathscr{C}^{\rm R}
\nonumber\\&
- (1-Z_\xi \xi_{\rm R})Z_g Z_A^{1/2} Z_C g_{\rm R} i  \mathscr{A}^\mu{}^{\rm R} \cdot (\partial_\mu \bar{\mathscr{C}}^{\rm R} \times \mathscr{C}^{\rm R})
\nonumber\\&
+ Z_\xi Z_g Z_A^{1/2} Z_C \xi_{\rm R} g_{\rm R} i \mathscr{A}^\mu{}^{\rm R} \cdot (\bar{\mathscr{C}}^{\rm R} \times \partial_\mu \mathscr{C}^{\rm R})
\nonumber\\&
+ {1 \over 2}Z_\lambda Z_\xi Z_g^2 Z_C^2 \lambda_{\rm R} \xi_{\rm R} (1-Z_\xi \xi_{\rm R}) g_{\rm R}^2 (i \mathscr{C}^{\rm R} \times \bar{\mathscr{C}}^{\rm R}) \cdot (i \mathscr{C}^{\rm R} \times \bar{\mathscr{C}}^{\rm R})
\nonumber\\&
 + {\lambda_{\rm R} \over 2} Z_\lambda \left( Z_C Z_A^{-1/2} \mathscr{B}^{\rm R} + Z_\lambda^{-1} Z_A^{1/2} \lambda_{\rm R}^{-1} \partial^\mu \mathscr{A}_\mu^{\rm R} - Z_\xi Z_g Z_C \xi_{\rm R} ig_{\rm R} \mathscr{C}^{\rm R} \times \bar{\mathscr{C}}^{\rm R}  \right)^2  .
\label{rbLag}
\end{align}
The total Lagrangian (\ref{rbLag}) is decomposed into a renormalization-factor independent part $\mathscr{L}_{\rm YM}^{\rm tot}{}^{\rm R}$ and the remaining part $\mathscr{L}_{\rm YM}^{\rm tot}{}^{\rm c}$ as
\begin{subequations}
\begin{align}
  \mathscr{L}_{\rm YM}^{\rm tot} =\,& \mathscr{L}_{\rm YM}^{\rm tot}{}^{\rm R} + \mathscr{L}_{\rm YM}^{\rm tot}{}^{\rm c} ,
\\
 \mathscr{L}_{\rm YM}^{\rm tot}{}^{\rm R} :=\,& 
 - {1 \over 4} (\partial_\mu \mathscr{A}_\nu^{\rm R} - \partial_\nu \mathscr{A}_\mu^{\rm R} + g_{\rm R} \mathscr{A}_\mu^{\rm R} \times \mathscr{A}_\nu^{\rm R} )^2
\nonumber\\&
 - {1 \over 2\lambda_{\rm R}} (\partial^\mu \mathscr{A}_\mu^{\rm R})^2
+ i \bar{\mathscr{C}}^{\rm R} \cdot \partial_\mu \partial^\mu \mathscr{C}^{\rm R}
\nonumber\\&
- (1-\xi_{\rm R}) g_{\rm R} i  \mathscr{A}^\mu{}^{\rm R} \cdot (\partial_\mu \bar{\mathscr{C}}^{\rm R} \times \mathscr{C}^{\rm R})
+  \xi_{\rm R} g_{\rm R} i \mathscr{A}^\mu{}^{\rm R} \cdot (\bar{\mathscr{C}}^{\rm R} \times \partial_\mu \mathscr{C}^{\rm R})
\nonumber\\&
+ {1 \over 2} \lambda_{\rm R} \xi_{\rm R} (1-\xi_{\rm R}) g_{\rm R}^2 (i \mathscr{C}^{\rm R} \times \bar{\mathscr{C}}^{\rm R}) \cdot (i \mathscr{C}^{\rm R} \times \bar{\mathscr{C}}^{\rm R})
\nonumber\\&
 + {\lambda_{\rm R} \over 2}  \left( \mathscr{B}^{\rm R} + \lambda_{\rm R}^{-1} \partial^\mu \mathscr{A}_\mu^{\rm R} -  \xi_{\rm R} ig_{\rm R} \mathscr{C}^{\rm R} \times \bar{\mathscr{C}}^{\rm R}  \right)^2  ,
 \label{rLag}
\\
 \mathscr{L}_{\rm YM}^{\rm tot}{}^{\rm c} :=\,& (\ref{rbLag}) - (\ref{rLag}) .
\end{align}
\end{subequations}
Here $\mathscr{L}_{\rm YM}^{\rm tot}{}^{\rm R}$ is obtained by setting all renormalization factors $Z \equiv 1$ in (\ref{rbLag}) and hence it is written in terms of the renormalized fields and renormalized parameters and has the same form as the bare Lagrangian $\mathscr{L}_{\rm YM}^{\rm tot}$, while $\mathscr{L}_{\rm YM}^{\rm tot}{}^{\rm c}$ is the counterterm defined by the difference $\mathscr{L}_{\rm YM}^{\rm tot} - \mathscr{L}_{\rm YM}^{\rm tot}{}^{\rm R}$.

\subsubsection{Renormalization of two-point functions}
\par
First, we calculate the vacuum polarization function of the gluon.
To the order $g^2$, there are three Feynman diagrams, see (a1), (a2) and (a3) in  Fig.\ref{fig:vacuum polarization}.

\unitlength=0.001in
\begin{figure}[htbp]
\begin{center}
\begin{picture}(4000,800)
\put(0,700){\mbox{(a1)}}%
\put(0,0){%
   \put(0,0){\includegraphics[height=1.9cm]{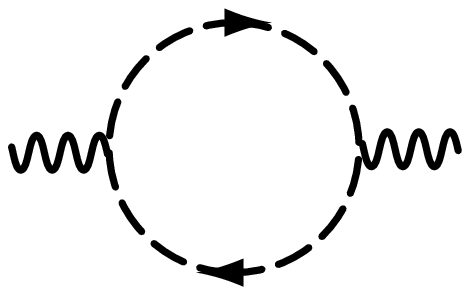}}%
   }%
\put(1500,700){\mbox{(a2)}}%
\put(1500,0){%
   \put(0,0){\includegraphics[height=2cm]{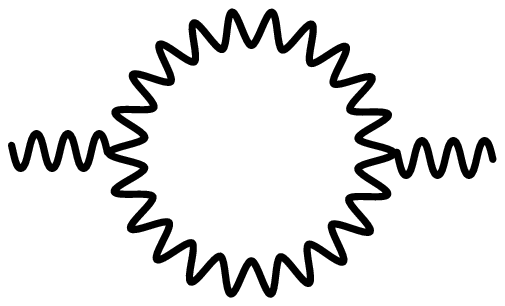}}%
   }%
\put(3000,700){\mbox{(a3)}}%
\put(3300,0){%
   \put(0,0){\includegraphics[height=2cm]{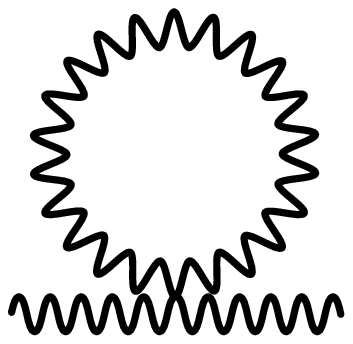}}%
   }%
\end{picture}
\caption{Vacuum polarization of the gluon.}
\label{fig:vacuum polarization}
\end{center}
\end{figure}

As a gauge-invariant regularization, we adopt the dimensional regularization.  Then we obtain the following result
($\epsilon :=2-D/2$).
\begin{subequations}
\begin{align}
\mbox{(a1)}
 &=C_2(G) \delta^{AB}\frac{(g\mu^{-\epsilon})^2}{(4\pi)^2}\frac i\epsilon
  \left[\frac1{12}q^2g_{\mu\nu}
        -\left\{\xi(1-\xi)-\frac16\right\}q_\mu q_\nu\right] ,
\\
\mbox{(a2)}
  &=\frac12C_2(G) \delta^{AB}
    \frac{(g\mu^{-\epsilon})^2}{(4\pi)^2}\frac i\epsilon
    \left\{\frac{19}6q^2g_{\mu\nu}
           -\frac{11}3q_\mu q_\nu
           +(1-\lambda)(q^2g_{\mu\nu}-q_\mu q_\nu)\right\} ,
\\
\mbox{(a3)} &=0 ,
\end{align}
\end{subequations}
where $C_2=C_2(G)$ is the quadratic Casimir operator in the adjoint representation of the gauge group $G$ defined by
$\delta^{AB} C_2(G)=f^{ACD}f^{BCD}$.
Hence the counterterms $\delta_{\rm T}$ and $\delta_{\rm L}$ for the transverse and longitudinal part of the vacuum polarization tensor are determined so as to satisfy the relation,
\begin{equation}
\mbox{(a1)}+\mbox{(a2)}+\mbox{(a3)}
 -i\delta_{\rm T}(q^2g_{\mu\nu}-q_\mu q_\nu) \delta^{AB}
 -i\frac{\delta_{\rm L}}\lambda q_\mu q_\nu \delta^{AB}
 \equiv0 ,
\end{equation}
which yields the result:
\begin{equation}
\delta_{\rm T}
 =\left(\frac{13}6-\frac\lambda2\right)
  \frac{(g\mu^{-\epsilon})^2}{(4\pi)^2}\frac{C_2(G)}{\epsilon} ,
\quad
\delta_{\rm L}
 =-\lambda\xi(1-\xi)
   \frac{(g\mu^{-\epsilon})^2}{(4\pi)^2}\frac{C_2(G)}{\epsilon} .
\end{equation}
On the other hand, the relationship
\begin{equation}
 \delta_{\rm T}=Z_A-1=Z_A^{(1)}+\cdots, \quad
 \delta_{\rm L}=Z_AZ_\lambda^{-1}-1=Z_A^{(1)}-Z_\lambda^{(1)}+\cdots ,
\end{equation}
must hold for the multiplicative renormalizability where we have defined the renormalization factor $Z$ order by order of the loop expansion, $Z=1+Z^{(1)}+Z^{(2)}+ \cdots$.
Thus we obtain the renormalization factors:
\begin{equation}
Z_A^{(1)} =\delta_{\rm T} 
 =\left(\frac{13}6-\frac\lambda2\right)
  \frac{(g\mu^{-\epsilon})^2}{(4\pi)^2} \frac{C_2(G)}{\epsilon} ,
\end{equation}
and
\begin{equation}
Z_\lambda^{(1)} =\delta_{\rm T}-\delta_{\rm L}
 =\left[\left(\frac{13}6-\frac\lambda2\right)+\lambda\xi(1-\xi)\right]
  \frac{(g\mu^{-\epsilon})^2}{(4\pi)^2} \frac{C_2(G)}{\epsilon} .
\end{equation}
Note that $\delta_{\rm T}$ and hence $Z_A$ is the same as in the FP case where the four ghost interaction does not exist.  
When $\xi \not=0,1$, however, we find that $\delta_{\rm L}\not=0$ or equivalently
$Z_A \not=Z_\lambda$.  On the contrary to the FP case, the longitudinal part of the gluon propagator must be renormalized in this case.

\begin{figure}[htbp]
\begin{center}
\begin{picture}(3100,800)
\put(0,650){\mbox{(b1)}}%
\put(0,0){%
   \put(0,0){\includegraphics[height=2cm]{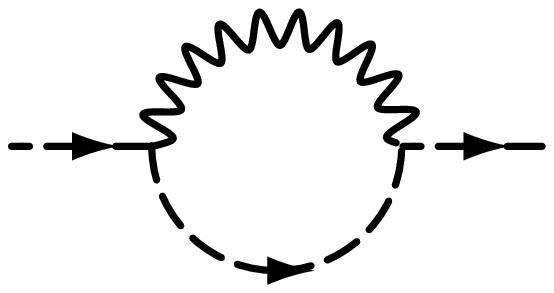}}%
   }%
\put(1600,650){\mbox{(b2)}}%
\put(1600,0){%
   \put(0,0){\includegraphics[height=2cm]{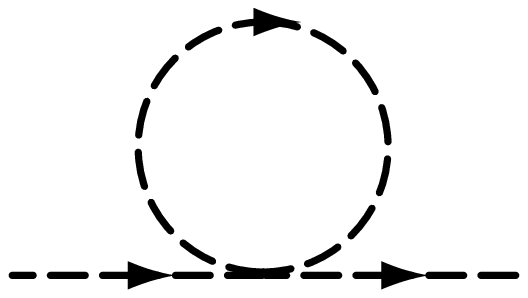}}%
   }%
\end{picture}
\caption{Vacuum polarization of the ghost}
\label{fig:ghost propagator}
\end{center}
\end{figure}

Next, the vacuum polarization function of the ghost is calculated in the similar way.
To the order $g^2$, there are two Feynman diagrams, see (b1) and (b2) in  Fig.\ref{fig:ghost propagator}.  
The explicit calculation shows that 
\begin{subequations}
\begin{align}
  \text{(b1)} &=\left(\frac{1}2+\frac{1-\lambda}{4}\right)
  \frac{(g\mu^{-\epsilon})^2}{(4\pi)^2}\frac{C_2(G)}{\epsilon} p^2 \delta^{AB} ,
\\
  \text{(b2)} &=0.
\end{align}
\end{subequations}
The counterterm $\delta_C$ is determined from 
\begin{equation}
 \text{(b1)}+\text{(b2)} - p^2 \delta^{AB} \delta_C =0 .
\end{equation}
Hence the counterterm 
$\delta_C =Z_C - 1 =Z_C^{(1)}+\cdots$ is equal to the renormalization constant $Z_C^{(1)}$:
\begin{equation}
Z_C^{(1)}
 =\delta_C
 =\frac{3-\lambda}4
  \frac{(g\mu^{-\epsilon})^2}{(4\pi)^2} \frac{C_2(G)}{\epsilon} .
\end{equation}
This is again the same as in the FP case.

\subsubsection{Renormalization of three-point function}

\begin{figure}[htbp]
\begin{center}
\begin{picture}(4000,1000)
\put(0,850){\mbox{(c1)}}%
\put(0,0){%
   \put(0,0){\includegraphics[height=2.5cm]{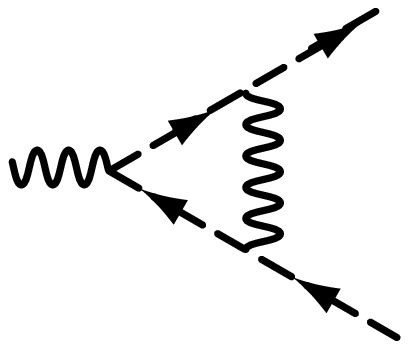}}%
   }%
\put(1400,850){\mbox{(c2)}}%
\put(1400,0){%
   \put(0,0){\includegraphics[height=2.5cm]{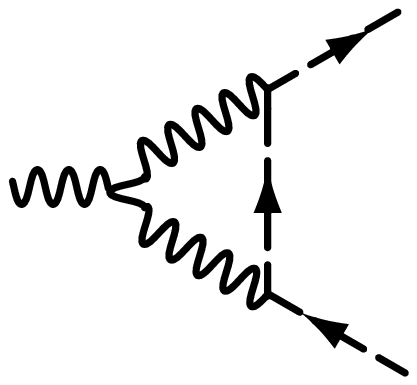}}%
   }%
\put(2800,850){\mbox{(c3)}}%
\put(2800,0){%
   \put(0,0){\includegraphics[height=2.5cm]{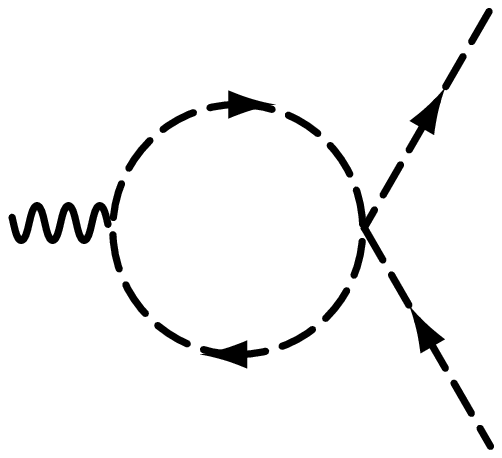}}%
   }%
\end{picture}
\caption{Radiative corrections for the gluon--ghost--anti-ghost vertex.}
\label{fig:3 point vertex}
\end{center}
\end{figure}

We consider the renormalization of three-point vertex.
For example, the Feynman diagrams for the radiative correction of the gluon--ghost--anti-ghost vertex to one-loop order is given in Fig.\ref{fig:3 point vertex}.

If we write the counterterm for the gluon--ghost--anti-ghost vertex function as
\begin{equation}
\begin{array}{c}
\begin{picture}(800,800)
   \put(40,0){\includegraphics[height=2cm]{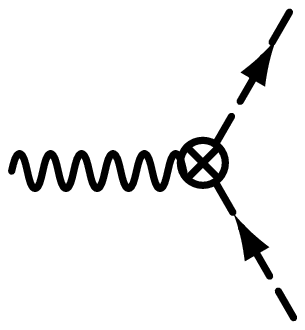}}%
   \put(570,700){\mbox{$p$}}%
   \put(700,150){\mbox{$q$}}%
   \put(100,500){\mbox{$C,\mu$}}%
   \put(700,600){\mbox{$A$}}%
   \put(500,30){\mbox{$B$}}%
\end{picture}
\end{array}
 =ig_{\rm R}f^{ABC} 
  \left[\xi_{\rm R}\delta_{AC\bar{C}}^1(p-q)
        -\delta_{AC\bar{C}}^2p\right]_\mu ,
\end{equation}
we find the renormalization factors are related as
\begin{align}
\delta_{AC\bar{C}}^1
 &=Z_CZ_A^{\frac12}Z_gZ_\xi-1
 =Z_C^{(1)}+\frac12Z_A^{(1)}+Z_g^{(1)}+Z_\xi^{(1)}
  +\cdots ,
\\
\delta_{AC\bar{C}}^2
  &=Z_CZ_A^{\frac12}Z_g-1
  =Z_C^{(1)}+\frac12Z_A^{(1)}+Z_g^{(1)}
   +\cdots .
\end{align}

\par
At $p=q$, the respective diagram is calculated as
\begin{subequations}
\begin{align}
\mbox{(c1)}_{p=q}
  &=-\frac12 C_2(G)f^{ABC}g^3
    \frac i{(4\pi)^2}\frac1\epsilon
    \frac\lambda4p^\mu ,
\label{eq:c1_p=q}
\\
\mbox{(c2)}_{p=q}
 &=-\frac12 C_2(G)f^{ABC}g^3\lambda
   \frac i{(4\pi)^2}\frac1\epsilon
   \frac34p_\mu ,
\label{eq:c2_p=q}
\\
\mbox{(c3)}_{p=q} &=0 .
\label{eq:c3_p=q}
\end{align}
\end{subequations}
By substituting 
(\ref{eq:c1_p=q}), (\ref{eq:c2_p=q}) and (\ref{eq:c3_p=q}) 
into
\begin{equation}
\mbox{(c1)}_{p=q}+\mbox{(c2)}_{p=q}+\mbox{(c3)}_{p=q}
 -igf^{ABC}\delta_{ACC}^2p_\mu
 \equiv0 ,
\end{equation}
it follows that
\begin{equation}
\delta_{AC\bar{C}}^2
 =-\frac12\lambda
    \frac{(g\mu^{-\epsilon})^2}{(4\pi)^2} \frac{C_2(G)}{\epsilon} .
\end{equation}
Hence the renormalization factor is obtained as
\begin{equation}
Z_g^{(1)}
 =\delta_{AC\bar{C}}^2-Z_C^{(1)}-\frac12Z_A^{(1)}
 =-\frac{11}6
    \frac{(g\mu^{-\epsilon})^2}{(4\pi)^2}\frac{C_2(G)}{\epsilon} .
\end{equation}

\par
At $p=0$, the respective diagram is calculated as
\begin{subequations}
\begin{align}
\mbox{(c1)}_{p=0}
  &=-\frac12 C_2(G)f^{ABC}g^3\xi\lambda
   \frac i{(4\pi)^2}\frac1\epsilon
   \left[(1-\xi)\left(\xi-\frac12\right)
         +\frac14\right]q^\mu ,
\label{eq:c1_p=0}
\\
\mbox{(c2)}_{p=0}
  &=-\frac12 C_2(G)f^{ABC}g^3\lambda\xi
    \frac i{(4\pi)^2}\frac1\epsilon
    \frac34q^\mu ,
\label{eq:c2_p=0}
\\
\mbox{(c3)}_{p=0}
 &=-\frac12 C_2(G)f^{ABC}g^3
     \lambda\xi(1-\xi)
     \frac i{(4\pi)^2}\frac1\epsilon
     \left(\xi-\frac12\right)q^\mu .
\label{eq:c3_p=0}
\end{align}
\end{subequations}
By substituting 
(\ref{eq:c1_p=0}), (\ref{eq:c2_p=0}) and (\ref{eq:c3_p=0})
into 
\begin{equation}
\mbox{(c1)}_{p=0}+\mbox{(c2)}_{p=0}+\mbox{(c3)}_{p=0}
 -igf^{ABC}\xi_{\rm R}\delta_{AC\bar{C}}^1q_\mu
 \equiv0 ,
\end{equation}
it follows that
\begin{equation}
\delta_{ACC}^1
  =\left[
    -\lambda(1-\xi)\left(\xi-\frac12\right)
    -\frac12\lambda
    \right]
    \frac{(g\mu^{-\epsilon})^2}{(4\pi)^2}\frac{C_2(G)}{\epsilon} .
\end{equation}
Then we obtain
\begin{align}
Z_\xi^{(1)}
 &=\delta_{AC\bar{C}}^1-\delta_{AC\bar{C}}^2
    \nonumber\\
  &=\lambda(\xi-1)\left(\xi-\frac12\right)
    \frac{(g\mu^{-\epsilon})^2}{(4\pi)^2}\frac{C_2(G)}{\epsilon} .
\end{align}
Accordingly, the renormalization constants of $\alpha$ and $\alpha'$ are obtained as
\begin{equation}
Z_\alpha^{(1)}
 =\left( \frac{13}6-\frac\alpha4  \right)
  \frac{(g\mu^{-\epsilon})^2}{(4\pi)^2} \frac{C_2(G)}{\epsilon} ,
\end{equation}
and
\begin{equation}
Z_{\alpha^\prime}^{(1)}
 =\left( \frac{13}6-\frac{\alpha+\alpha'}{2} \right)
  \frac{(g\mu^{-\epsilon})^2}{(4\pi)^2} \frac{C_2(G)}{\epsilon} .
\end{equation}

\section{Renormalization group flow and fixed points}

 Using the above result, the renormalization group (RG) functions are obtained as follows.
The $\beta$-function is obtained as
\begin{equation}
  \beta(g_{\rm R}) :=\mu {\partial g_{\rm R} \over \partial \mu} 
=- g_{\rm R} \mu {\partial \over \partial \mu} \ln Z_g 
\cong - g_{\rm R} \mu {\partial \over \partial \mu}  Z_g^{(1)} .
\end{equation}
It turns out that the $\beta$-function does not depend on the gauge parameters $\lambda$ and $\xi$:
\begin{equation}
  \beta(g_{\rm R}) :=\mu {\partial g_{\rm R} \over \partial \mu}
=- {1 \over 16\pi^2} {11 \over 3}C_2(G) g_{\rm R}^3  .
\end{equation}
 Similarly, we obtain the RG functions:
\begin{equation}
\gamma_\xi :=\mu\frac\partial{\partial\mu} \xi_{\rm R}
 =2 \lambda_{\rm R} \xi_{\rm R}(\xi_{\rm R}-1) \left(\xi_{\rm R}-{1 \over 2}\right)
  \frac{C_2(G)}{(4\pi)^2} g_{\rm R}^2 ,
\end{equation}
and
\begin{equation}
\gamma_\lambda
:=\mu\frac\partial{\partial\mu} \lambda_{\rm R}
 =2\lambda_{\rm R} \left[ \frac{13}6-\frac{\lambda_{\rm R}}{2} + \lambda_{\rm R}\xi_{\rm R}(1-\xi_{\rm R})\right]
  \frac{C_2(G)}{(4\pi)^2} g_{\rm R}^2  .
\end{equation}

\par
The RG flow in three-dimensional parameter space $(\xi,\lambda,g)$ is determined by solving simultaneous differential equations:
\begin{subequations}
\begin{align}
  \mu\frac{\partial \xi}{\partial\mu}  &=2\lambda\xi(\xi-1) \left(\xi-{1 \over 2}\right)
  \frac{C_2(G)g^2}{(4\pi)^2}  ,
\label{DEa}
  \\
  \mu\frac{\partial \lambda}{\partial\mu}  &=2\lambda \left[ \frac{13}6-\frac\lambda2 + \lambda\xi(1-\xi)\right]
  \frac{C_2(G)g^2}{(4\pi)^2}  ,
\label{DEb}
\\
   \mu {\partial g \over \partial \mu} 
 &=-  {11 \over 3}{C_2(G) g^3 \over (4\pi)^2}  ,
\label{DEc}
\end{align}
\end{subequations}
where we have omitted the subscript ${\rm R}$ for the renormalized quantity.
\par
As is well known, the equation (\ref{DEc}) is solved exactly, 
\begin{equation}
  g^2(\mu) = {g^2(\mu_0) \over 1+ {22 \over 3}{C_2(G) \over (4\pi)^2}g^2(\mu_0) \ln {\mu \over \mu_0}} 
  = {1 \over {22 \over 3}{C_2(G) \over (4\pi)^2} \ln {\mu \over \Lambda_{\rm QCD}}} ,
\end{equation}
where we have used the boundary condition $g(\mu_0)=\infty$ at $\mu_0=\Lambda_{\rm QCD}$. 
The remaining two equations (\ref{DEa}) and (\ref{DEb}) can not be solved exactly.  

\subsection{Fixed points}

\par
First, we obtain the fixed point of the RG.
Note that the derivative 
${1 \over g^2} \mu {\partial \over \partial \mu}$
in (\ref{DEa}), (\ref{DEb}) is rewritten as
\begin{equation}
  {1 \over g^2} \mu {\partial \over \partial \mu}
  = {22 \over 3}{C_2(G) \over (4\pi)^2} \ln {\mu \over \Lambda_{\rm QCD}}  \mu {\partial \over \partial \mu}
  =  {22 \over 3}{C_2(G) \over (4\pi)^2}   {\partial \over \partial \ln \ln {\mu \over \Lambda_{\rm QCD}}} .
\end{equation}
Then the fixed point (to one-loop order) is obtained by solving the algebraic equation simultaneously: 
\begin{equation}
   \lambda\xi(\xi-1) \left(\xi-{1 \over 2}\right)  =0 ,
  \quad 
   \lambda \left[ \frac{13}6-\frac\lambda2 + \lambda\xi(1-\xi)\right]
 =0 .
\end{equation}
We find one line of fixed points and three isolated fixed points in the $(\xi,\lambda)$ plane or equivalently four isolated fixed points in the $(\alpha, \alpha')$ plane:
\begin{enumerate}
\item[A.] 
 The line of fixed points: $\lambda =0, \xi \in {\bf R}$ corresponds to an isolated fixed point $(\alpha, \alpha')=(0,0)$.
\item[B.]
 $(\xi,\lambda) =({1 \over 2},{26 \over 3})$ corresponds to $(\alpha, \alpha')=({26 \over 3},0)$.
\item[C.]
 $(\xi,\lambda) =(0,{13 \over 3})$ corresponds to $(\alpha, \alpha')=(0,{13 \over 3})$.
\item[D.]
 $(\xi,\lambda) =(1,{13 \over 3})$ corresponds to $(\alpha, \alpha')=({26 \over 3},-{13 \over 3})$.
\end{enumerate}
If the two parameters $\xi, \lambda$ are set equal to one of the fixed points, the theory remains forever on the fixed.
If the system starts from other points and the scale $\mu$ is decreased, it evolves into the infrared (IR) region according to a couple of differential equations (\ref{DEa})--(\ref{DEc}).

\subsection{RG flow in the neighborhood of fixed points}

\begin{table}[bp]
\[
\begin{array}{|c||c|c||cc|c|}\hline
&
\multicolumn{2}{|l||}{\mbox{Eigenvalue}} &
\multicolumn{3}{|l|}{\mbox{Eigenvector}}
\\
&
\multicolumn{2}{|l||}
{\left( \frac{13}{3}\frac{g^2C_2(G)}{(4\pi)^2}\times \right)}&
(\xi,\lambda) &
\multicolumn{2}{l|}{(\alpha,\alpha^\prime)}
\\ \hline\hline
\mbox{A} &
1&
\mbox{IR fixed point}&
&
(1,a) &
\mbox{any lines}
\\ \hline
\mbox{B} &
-1&
\mbox{UV fixed point} &
(1,a) &
(1,a) &
\mbox{any lines}
\\ \hline
\mbox{C} &
1  &
\multicolumn{1}{|l||}{\mbox{~~~Saddle}}&
(13,3) &
(1,-2) &
\mbox{line $\rm I\!V$}
 \\
&
-1&
\multicolumn{1}{|r||}{\mbox{point~~~}}  &
(0,1) &
(0,1) &
\mbox{line $\rm I\!I$}
\\ \hline
\mbox{D} &
1 &
\multicolumn{1}{|l||}{\mbox{~~~Saddle}}&
(-13,3) &
(0,1) &
\mbox{line $\rm V$}
\\
&
-1 &
\multicolumn{1}{|r||}{\mbox{point~~~}} &
(1,-2) &
(0,1) &
\mbox{line $\rm I\!I\!I$}
\\ \hline
\end{array}
\]
\caption{Eigenvalues and eigenvectors of the linearized RG equation where the lines, ${\rm I\!I}$, ${\rm I\!I\!I}$, ${\rm I\!V}$ are defined below.  At the IR fixed point A and UV fixed point B, two eigenvalues are degenerate.  
}
\label{Table1}
\end{table}
\par
In the neighborhood of the respective fixed point $(X_1^*, X_2^*)$ in the plane $(X_1,X_2)=(\xi, \lambda)$ or $(\alpha, \alpha')$, we can study the behavior of the RG flow analytically.
By taking into account only the terms which are linear in the
infinitesimal deviation $\delta X_1 :=X_1-X_1^*, \delta X_2 :=X_2-X_2^*$ from the fixed point, a set of RG
equations, (\ref{DEa}) and (\ref{DEb}), is reduced to the form:
$
\left(\begin{array}{c}\gamma_{X_1} \\ \gamma_{X_2} \\ \end{array}\right)
\sim {\bf A}
\left(\begin{array}{c}\delta X_1 \\ \delta X_2 \\ \end{array} \right) ,
$
where ${\bf A}$ is a two by two matrix.
\par
In $(\xi,\lambda)$ plane, the set of linearized RG equations reads
\begin{subequations}
\begin{align}
\mbox{B}\left(\frac{1}{2},\frac{26}{3}\right) :&
\left(\begin{array}{c}\gamma_\xi \\ \gamma_\lambda \\ \end{array}\right)
\sim
-\frac{13}{3}\frac{g^2C_2(G)}{(4\pi)^2}
\left(\begin{array}{cc}1 & 0 \\ 0 & 1 \\ \end{array}\right)
\left(\begin{array}{c}\delta\xi \\ \delta\lambda \\ \end{array}\right) .
 \\
\mbox{C}\left(0,\frac{13}{3}\right) :&
\left(\begin{array}{c}\gamma_\xi \\ \gamma_\lambda \\ \end{array}\right)
\sim
\frac{13}{3}\frac{g^2C_2(G)}{(4\pi)^2}
\left(\begin{array}{cc}1 & 0 \\ \frac{26}{3} & -1 \\ \end{array}\right)
\left(\begin{array}{c}\delta\xi \\ \delta\lambda \\ \end{array}\right) .
 \\
\mbox{D}\left(1,\frac{13}{3}\right) :&
\left(\begin{array}{c}\gamma_\xi \\ \gamma_\lambda \\ \end{array}\right)
\sim
\frac{13}{3}\frac{g^2C_2(G)}{(4\pi)^2}
\left(\begin{array}{cc}1 & 0 \\ -\frac{26}{3} & -1 \\ \end{array}\right)
\left(\begin{array}{c}\delta\xi \\ \delta\lambda \\ \end{array}\right)
.
\end{align}
\end{subequations}
Similarly in $(\alpha,\alpha^\prime)$ plane, we obtain
\begin{subequations}
\begin{align}
\mbox{A}\left(0,0\right)  :&
\left(\begin{array}{c}\gamma_\alpha \\ \gamma_{\alpha^\prime} \
 \end{array}\right)
\sim
\frac{13}{3}\frac{g^2C_2(G)}{(4\pi)^2}
\left(\begin{array}{cc}1 & 0 \\ 0 & 1 \\ \end{array}\right)
\left(\begin{array}{c}\delta\alpha \\ \delta\alpha^\prime \
 \end{array}\right) .
\\
\mbox{B}\left(\frac{26}{3},0\right)
 :&
\left(
\begin{array}{c}\gamma_\alpha \\ \gamma_{\alpha^\prime} \\ \end{array}
\right)
\sim
-\frac{13}{3}\frac{g^2C_2(G)}{(4\pi)^2}
\left(\begin{array}{cc}1 & 0 \\ 0 & 1 \\ \end{array}\right)
\left(\begin{array}{c}\delta\alpha \\ \delta\alpha^\prime \
 \end{array}\right) .
\\
\mbox{C}\left(0,\frac{13}{3}\right)
 :&
\left(\begin{array}{c}\gamma_\alpha \\ \gamma_{\alpha^\prime} \
 \end{array}\right)
\sim
\frac{13}{3}\frac{g^2C_2(G)}{(4\pi)^2}
\left(\begin{array}{cc}1 & 0 \\ -1 & -1 \\ \end{array}\right)
\left(\begin{array}{c}\delta\alpha \\ \delta\alpha^\prime \
 \end{array}\right) .
\\
\mbox{D}\left(-\frac{13}{3},\frac{26}{3}\right)
 :&
\left(\begin{array}{c}\gamma_\alpha \\ \gamma_{\alpha^\prime} \
 \end{array}\right)
\sim
\frac{13}{3}\frac{g^2C_2(G)}{(4\pi)^2}
\left(\begin{array}{cc}-1 & 0 \\ 1 & 1 \\ \end{array}\right)
\left(\begin{array}{c}\delta\alpha \\ \delta\alpha^\prime \
 \end{array}\right) .
\end{align}
\end{subequations}
The respective matrix characterizing the behavior of the RG flow in the neighborhood of the respective fixed point has the eigenvalue and the corresponding eigenvector enumerated in Table \ref{Table1}.
The direction of the flow is determined at the respective fixed point.  We will see that these results are consistent with the global flow diagram given in Fig.\ref{fig:RGflow} below.

\subsection{Global behavior of the RG flow}

\par
We find that $\xi \equiv 0$, $\xi \equiv 1/2$, and $\xi \equiv 1$ are solutions of the equation (\ref{DEa}).  This implies that the RG flow starting from the point on one of the three planes,  
$(0,\lambda,g)$, $(1/2,\lambda,g)$, $(1,\lambda,g)$, is always kept on the respective plane. 
On the three planes, moreover, the equation (\ref{DEb}) can be solved exactly. 
On the plane $(1/2,\lambda,g)$, the RG flow
in the region  
$0<\lambda<\frac{26}{3}$
 obeys 
\begin{equation}
 \lambda(\mu) = \frac{26}{3}
\left\{
1 + C \left(\ln \frac{\mu}{\Lambda_{\rm QCD}} \right)^{-\frac{13}{22}}  \right\}^{-1} ,
\end{equation}
where $C$ is a positive constant.
We see that 
$\lambda$ approaches to the ultraviolet (UV) fixed point $\lambda \uparrow \frac{26}{3}$ in the UV limit $\mu \uparrow \infty$, while 
$\lambda \downarrow 0$ monotonically as 
$\mu \downarrow \Lambda_{\rm QCD}$. 
On the other hand, the RG flow in the region $\lambda>\frac{26}{3}$ is described by
\begin{equation}
 \lambda(\mu) = \frac{26}{3}
\left\{
1 - C \left(\ln \frac{\mu}{\Lambda_{\rm QCD}} \right)^{-\frac{13}{22}}  \right\}^{-1} ,
\end{equation}
where 
$\lambda$ approaches to the UV fixed point $\lambda \uparrow \frac{26}{3}$ in the UV limit $\mu \uparrow \infty$, while 
$\lambda \uparrow \infty$ monotonically as 
$\mu \downarrow \Lambda_{\rm QCD}$. 
By substituting 
$
 \ln \frac{\mu}{\Lambda_{\rm QCD}}
= \left\{\frac{22}{3}\frac{C_2(G)}{(4\pi)^2}g^2\right\}^{-1}
$
into the above equation, 
the equation of the RG flow on the plane $(1/2,\lambda,g)$ is obtained 
\begin{equation}
 \lambda
= \frac{26}{3}
\left\{
1 \pm C \left(\frac{22}{3}\frac{C_2(G)}{(4\pi)^2}g^2\right)^{\frac{13}{22}}
  \right\}^{-1} .
\end{equation}
The RG flows on the plane $(0,\lambda,g)$ and $(1,\lambda,g)$ are governed by the same equations which are obtained by replacing $26/3$ with $13/3$.  

\par
\begin{figure}[htbp]
\begin{center}
\begin{picture}(5100,2600)(0,0)
\put(0,2400){\mbox{(a)}}%
\put(0,0){%
   \put(0,0){\includegraphics[height=6cm]{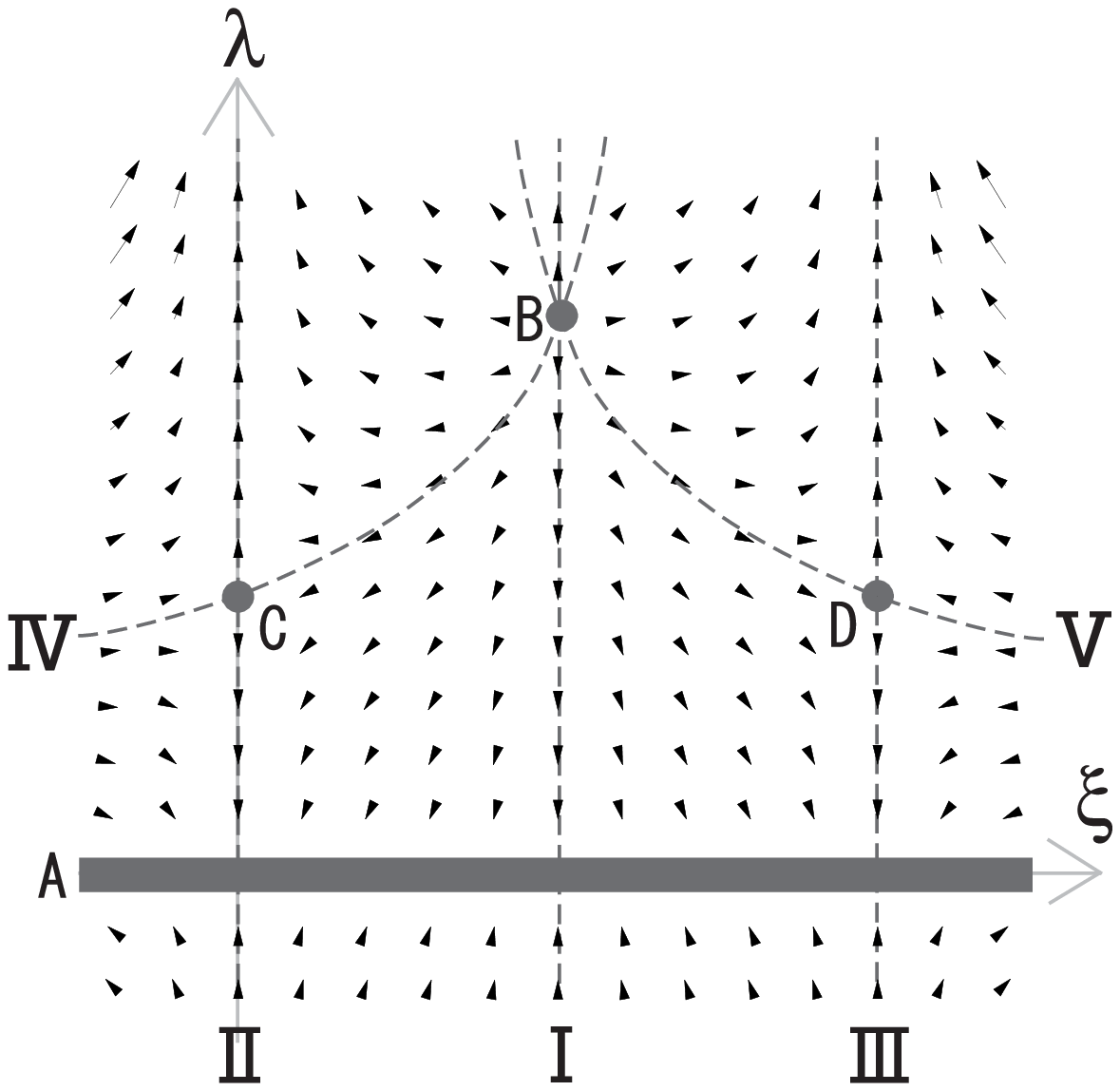}}%
   }%

\put(2800,2400){\mbox{(b)}}%
\put(2800,0){%
   \put(0,0){\includegraphics[height=6cm]{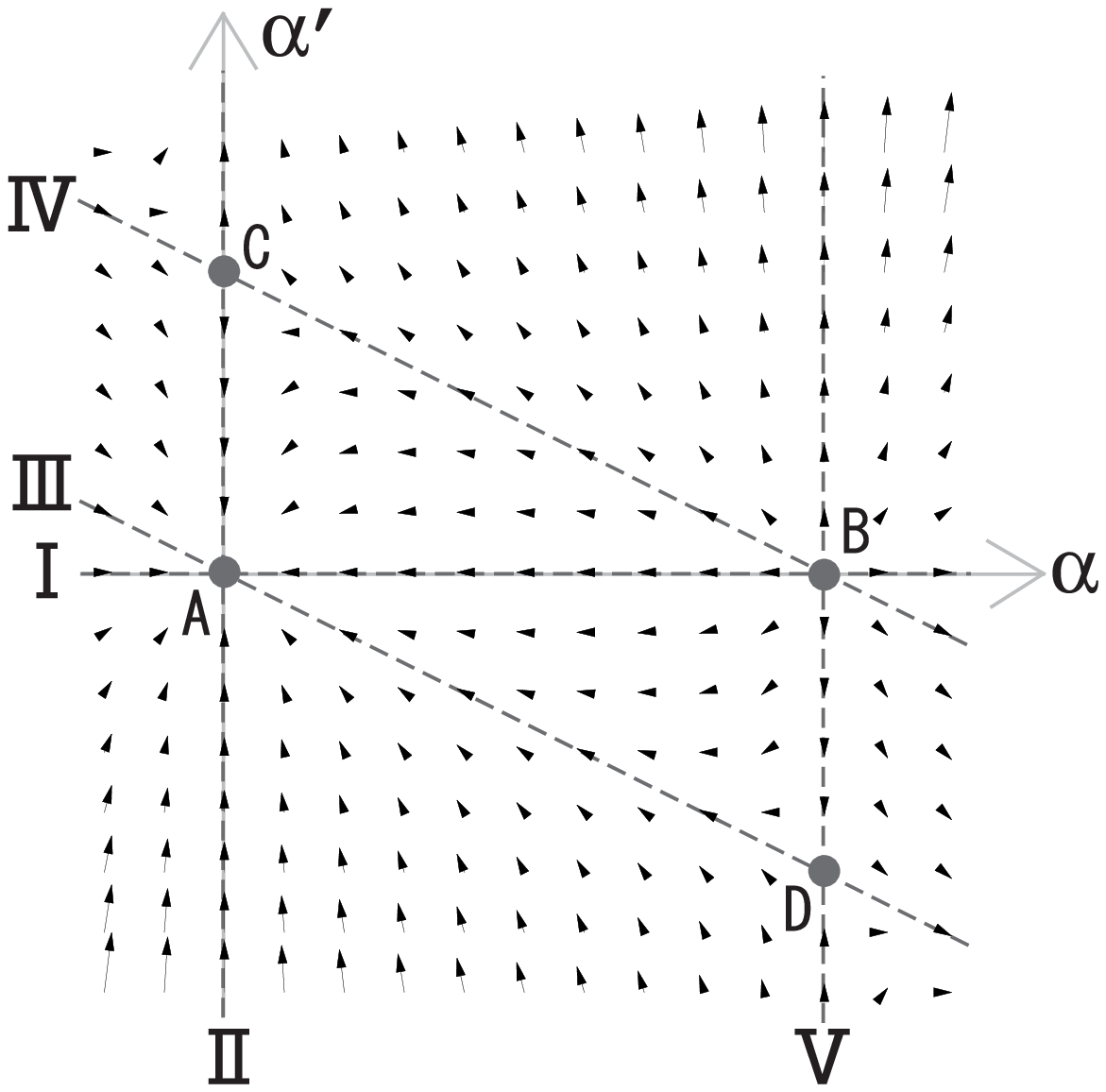}}%
   }%
%
\end{picture}
\caption{RG flows in  $(\xi,\lambda)$ plane (a) and in  $(\alpha, \alpha')$ plane (b).}
\label{fig:RGflow}
\end{center}
\end{figure}

 
The global behavior of the RG flow is obtained by solving (\ref{DEc})--(\ref{DEb}) numerically. 
In Fig.\ref{fig:RGflow}, the RG flow is drawn on the plane $(\xi,\lambda)$ and the plane $(\alpha, \alpha')$.  The direction of the arrow denotes the direction towards the IR region and the length of the arrow is proportional to the magnitude of the vector $\mu {d \over d\mu} (\xi, \lambda)/g^2$.  In the neighborhood of the respective fixed point, we see that the numerical result agrees with the analytical result given in Table \ref{Table1} of the previous subsection.  

Among the RG flows, the five RG flows
(I, ${\rm I\!I}$, ${\rm I\!I\!I}$, ${\rm I\!V}$, V) connecting the fixed points $A,B,C,D$ form the watershed (or backbone) in the flow diagram.
\begin{subequations}
\begin{alignat}{3}
 \text{I.} & \quad & \quad \xi & ={1 \over 2} , & \quad & \alpha' =0.
 \\
 \text{${\rm I\!I}$.} & & \quad \xi & =0 , & & \alpha =0.
 \\
 \text{${\rm I\!I\!I}$.} & & \quad \xi & =1, & & \alpha' =- {1 \over 2} \alpha.
 \\
 \text{${\rm I\!V}$.} & & \quad \lambda & ={13 \over 3}{1 \over 1-\xi}, & & \alpha' =- {1 \over 2} \alpha + {13 \over 3}.
 \\
 \text{V.} & & \quad \lambda & ={13 \over 3}{1 \over \xi}, & & \alpha ={26 \over 3}.
\end{alignat}
\end{subequations}

Since the flow is symmetric for the reflection with respect to the straight line I:$\xi=1/2$, we focus on the region $\xi \le 1/2$.
The flow starting from the initial position below ${\rm I\!V}$ runs towards the line $A$ of fixed points and eventually arrive at $A$.  
If it arrive at a fixed point on $A$  with a certain value of $\xi$ depending on the initial position, then it does not move anymore.  
On the other hand, the flow starting from the initial position above ${\rm I\!V}$ runs away into the infinity, $\lambda=+\infty$.  
Here the flow on the line I and ${\rm I\!I}$ is not an exception. 
However, it should be remarked that the fixed point $B$ is IR repulsive in both directions, while the fixed point $C$ is IR attractive on ${\rm I\!V}$ and repulsive on ${\rm I\!I}$.
In view of these, it turns out that any fixed point on $A$ is IR  stable, while the fixed point $B$ on I is a rather special fixed point which is IR unstable (UV stable).%
\footnote{
This does not imply that the similar result is obtained also for the MA gauge.  For example, $\alpha=0$ is not a fixed point in the MA gauge.  See \cite{KIMS01} for details.
}
\par
We have shown that the three fixed points $B,C,D$ for the gauge parameter $\xi,\lambda$ are located on the line I, ${\rm I\!I}$, ${\rm I\!I\!I}$ ($\xi=1/2,0,1$), respectively.  
On the lines I, ${\rm I\!I}$, ${\rm I\!I\!I}$, the RG flow is confined in the respective line, the Lagrangian takes the following form.

\begin{itemize}
\item[I.] 
 $\xi=\frac12$ (i.e., $\alpha \in {\bf R}$, $\alpha'=0$):
      The GF+FP term is invariant under the FP ghost conjugation and the orthosymplectic transformation $OSp(4|2)$ \cite{KondoII}.
\begin{equation}
  \mathscr{L}_{\rm GF+FP} =i \bm{\delta}_{\rm B} \bar{\bm{\delta}}_{\rm B} \left( {1 \over 2} \mathscr{A}_\mu \cdot \mathscr{A}^\mu  -{\alpha \over 2}i \mathscr{C} \cdot \bar{\mathscr{C}} \right) .
\end{equation}
There is a four-ghost interaction.

\item[${\rm I\!I}$.] 
 $\xi=0$ (i.e., $\alpha=0$, $\alpha' \in {\bf R}$):
      The GF+FP term is invariant under the global shift of anti-ghost $\bar{\mathscr{C}}$:
\begin{align}
 \mathscr{L}_{\rm GF+FP} &=
  {\alpha' \over 2} \mathscr{B} \cdot \mathscr{B} 
+ \mathscr{B} \cdot \partial_\mu \mathscr{A}^\mu + i \bar{\mathscr{C}} \cdot \partial_\mu \mathscr{D}^\mu[\mathscr{A}]\mathscr{C} .
\end{align}
There is no 4-ghost interaction.  This Lagrangian is the same as that in the conventional Lorentz gauge.  

\item[${\rm I\!I\!I}$.]
  $\xi=1$ (i.e., $\alpha'=-{1 \over 2}\alpha$):
      The GF+FP term is invariant under the global shift of ghost $\mathscr{C}$:
\begin{align}
 \mathscr{L}_{\rm GF+FP} 
&=
{\lambda \over 2} \mathscr{B} \cdot \mathscr{B} 
+ \mathscr{B} \cdot \partial_\mu \mathscr{A}^\mu
+  i \bar{\mathscr{C}} \cdot \mathscr{D}^\mu[\mathscr{A}] \partial_\mu \mathscr{C} .
\end{align}
There is no four-ghost interaction.  

\end{itemize}
The choice ${\rm I\!I}$ or ${\rm I\!I\!I}$ eliminates the four ghost interaction and the Yang-Mills theory reduces to the FP case. 
Once $\xi=0$ or $\xi=1$ is chosen,  $\xi$ is not renormalized by quantum corrections, since $\xi=0$ and $\xi=1$ are fixed point of the renormalization group.  
Then  the FP Lagrangian is preserved under the renormalization.  
\par
In ${\rm I\!I}$ and ${\rm I\!I\!I}$, the role of ghost and anti-ghost is interchanged.
The FP ghost conjugation invariance is broken in the usual FP Lagrangian where the ghost and anti-ghost are not treated on equal footing (except for the Landau gauge).
In other words, the FP ghost conjugation invariance is recovered for $\alpha'=0$ (i.e., $\xi=1/2$ or $\lambda=\alpha$) by including the quartic ghost interaction even for $\alpha=0$.

\par
We must keep in mind that these results are obtained to one-loop order.  Therefore, the details of the flow diagram may change if we include higher-order corrections.  
The higher-order result is not known to date and will be given elsewhere.
Nevertheless, the existence of the fixed point at $\lambda=0$ remains true to any finite order of perturbation.  The existence of the lines, I, ${\rm I\!I}$ and ${\rm I\!I\!I}$ are also guaranteed even after the inclusion of higher order terms, since it is protected by the symmetry dictated in the above.  This is because the symmetry can not be broken as far as the perturbation series to all orders are not summed up.

\section{Renormalizing the composite operator of mass dimension 2}

In this section we discuss the renormalization of the composite operator of mass dimension 2 and its BRST and anti-BRST invariance under the renormalization.

\subsection{On-shell BRST transformation}

\par
By eliminating the Nakanishi-Lautrup field $\mathscr{B}$, the on-shell BRST and anti-BRST transformations are obtained as
\begin{align}
 \bm{\delta}_{\rm B} \bar{\mathscr{C}}(x) 
 &=i \left[ - {1 \over \lambda} \partial^\mu \mathscr{A}_\mu(x) + \xi ig \mathscr{C}(x) \times \bar{\mathscr{C}}(x)  \right] ,
\\
 \bar{\bm{\delta}}_{\rm B} \mathscr{C}(x) 
&=i \left[  {1 \over \lambda} \partial^\mu \mathscr{A}_\mu(x) - (\xi-1) ig \mathscr{C}(x) \times \bar{\mathscr{C}}(x)  \right]  .
\label{BRSTonshell}
\end{align}

The nilpotency of the on-shell transformation is partially broken%
\footnote{
An elegant proof of the unitarity of the gauge theory is given based on the nilpotency of the BRST transformation, see e.g. \cite{Kugo89}.
The nilpotency is indeed broken in the on-shell BRST transformation which is obtained by eliminating the NL field.
However, the nilpotency is not the only way to show the unitarity.
Even in this case, it is possible to show the unitarity order by order of the perturbation theory based on the Feynman diagrams without the NL fields.
}
by the equation of motion of ghost and anti-ghost:
\begin{subequations}
\begin{align}
 (\bm{\delta}_{\rm B})^2 \mathscr{A}_\mu(x) &=0  \\
 (\bm{\delta}_{\rm B})^2 \mathscr{C}(x) &=0 ,
\\
 (\bm{\delta}_{\rm B})^2 \bar{\mathscr{C}}(x) &={-i \over \lambda} {\delta \mathscr{L}_{\rm YM}^{\rm tot} \over \delta \bar{\mathscr{C}}} 
 \nonumber\\
 &=\partial^\mu \mathscr{D}_\mu \mathscr{C} - g\xi (\partial^\mu \mathscr{A}_\mu \times \mathscr{C}) + ig^2 \lambda \xi (\xi-1) (\mathscr{C} \times \bar{\mathscr{C}}) \times \mathscr{C} ,
\label{BRSTnA3}
\end{align}
\end{subequations}
and
\begin{subequations}
\begin{align}
 (\bar{\bm{\delta}}_{\rm B})^2 \mathscr{A}_\mu(x) &=0 , 
 \\
 (\bar{\bm{\delta}}_{\rm B})^2 \mathscr{C}(x) &={-i \over \lambda} {\delta \mathscr{L}_{\rm YM}^{\rm tot} \over \delta \mathscr{C}}  
  \nonumber\\
 &=\partial^\mu \mathscr{D}_\mu \bar{\mathscr{C}} - g\xi (\partial^\mu \mathscr{A}_\mu \times \bar{\mathscr{C}}) - ig^2 \lambda \xi (\xi-1) (\mathscr{C} \times \bar{\mathscr{C}}) \times \bar{\mathscr{C}} ,
\\
 (\bar{\bm{\delta}}_{\rm B})^2 \bar{\mathscr{C}}(x) &=0 .
\label{BRSTnA4}
\end{align}
\end{subequations}
Moreover, the anti-commutatibity is also broken in the similar way:
\begin{subequations}
\begin{align}
 (\bm{\delta}_{\rm B} \bar{\bm{\delta}}_{\rm B} + \bar{\bm{\delta}}_{\rm B} \bm{\delta}_{\rm B} ) \mathscr{A}_\mu(x) &=0 , 
 \\
 (\bm{\delta}_{\rm B} \bar{\bm{\delta}}_{\rm B} + \bar{\bm{\delta}}_{\rm B} \bm{\delta}_{\rm B} ) \mathscr{C}(x) &={1 \over \lambda} {\delta \mathscr{L}_{\rm YM}^{\rm tot} \over \delta \bar{\mathscr{C}}} ,
 \\
 (\bm{\delta}_{\rm B} \bar{\bm{\delta}}_{\rm B} + \bar{\bm{\delta}}_{\rm B} \bm{\delta}_{\rm B} ) \bar{\mathscr{C}}(x) &={1 \over \lambda} {\delta \mathscr{L}_{\rm YM}^{\rm tot} \over \delta \mathscr{C}} .
\label{BRSTnA5}
\end{align}
\end{subequations}

\subsection{Composite operator of mass dimension 2}

We define the composite operator $\mathcal{O}$ as a linear combination of two composite operators of mass dimension 2:
\begin{equation}
  \mathcal{O} =(\Omega^{(D)})^{-1} \int d^D x \  \left[ 
 {1 \over 2} \mathscr{A}_\mu(x) \cdot \mathscr{A}^\mu{}(x) + \lambda i \bar{\mathscr{C}}(x) \cdot \mathscr{C}(x)    \right] .
 \label{combi2}
\end{equation}
The on-shell BRST transformation of the operator $\mathcal{O}$ is calculated as
\begin{align}
  \bm{\delta}_{\rm B} \mathcal{O} &=(\Omega^{(D)})^{-1} \int d^D x \  \bm{\delta}_{\rm B} \left[ 
 {1 \over 2} \mathscr{A}_\mu(x) \cdot \mathscr{A}^\mu{}(x) + \lambda i \bar{\mathscr{C}}(x) \cdot \mathscr{C}(x)    \right] 
\nonumber\\
 &=(\Omega^{(D)})^{-1} \int d^D x \   \left[ 
  \mathscr{A}_\mu(x) \cdot \bm{\delta}_{\rm B} \mathscr{A}^\mu(x) - \lambda i \bar{\mathscr{C}}(x) \cdot \bm{\delta}_{\rm B} \mathscr{C}(x)   
+ \lambda i  \bm{\delta}_{\rm B} \bar{\mathscr{C}}(x) \cdot \mathscr{C}(x)    \right] 
\nonumber\\
 &=(\Omega^{(D)})^{-1} \int d^D x \   \Big[ 
  \mathscr{A}_\mu(x) \cdot \partial^\mu \mathscr{C}(x) + \lambda i \bar{\mathscr{C}}(x) \cdot {g \over 2}(\mathscr{C}(x) \times \mathscr{C}(x))  
\nonumber\\
& \quad \quad \quad \quad + \partial^\mu \mathscr{A}_\mu(x) \cdot \mathscr{C}(x) - \lambda \xi i g (\mathscr{C}(x) \times \bar{\mathscr{C}}(x)) \cdot \mathscr{C}(x)    \Big] 
\nonumber\\
&=(\Omega^{(D)})^{-1} \int d^D x \left\{ \partial^\mu [\mathscr{A}_\mu(x) \cdot \mathscr{C}(x)] 
 + \lambda \left( {1 \over 2}-\xi \right) i \bar{\mathscr{C}}(x) \cdot g(\mathscr{C}(x) \times \mathscr{C}(x))  
\right\} .
\end{align}
In the similar way, the on-shell anti-BRST transformation of the operator $\mathcal{O}$ is calculated as
\begin{equation}
  \bar{\bm{\delta}}_{\rm B} \mathcal{O} =(\Omega^{(D)})^{-1} \int d^D x \left\{ \partial^\mu [\mathscr{A}_\mu(x) \cdot \bar{\mathscr{C}}(x)] 
 + \lambda \left( {1 \over 2}-\xi \right) i \bar{\mathscr{C}}(x) \cdot g(\bar{\mathscr{C}}(x) \times \mathscr{C}(x))  
\right\} .
\end{equation}
Therefore, the composite operator $\mathcal{O}$ is invariant under the BRST and anti-BRST transformations when 
\begin{equation}
 \xi= {1 \over 2} \quad \text{or} \quad \lambda=0 ,
\end{equation}
i.e., on the line I and A in the $(\xi,\lambda)$ plane, or on the line I in the $(\alpha, \alpha')$ plane.
For $\xi=1/2$, the on-shell BRST and anti-BRST transformation reads 
\begin{align}
 \bm{\delta}_{\rm B} \bar{\mathscr{C}}(x) 
 &=-{i \over \alpha} \partial^\mu \mathscr{A}_\mu(x) - {1 \over 2}g \mathscr{C}(x) \times \bar{\mathscr{C}}(x)   ,
\\
 \bar{\bm{\delta}}_{\rm B} \mathscr{C}(x) 
&=+{i \over \alpha} \partial^\mu \mathscr{A}_\mu(x) - {1 \over 2}g \mathscr{C}(x) \times \bar{\mathscr{C}}(x)    .
\label{BRSTonshell2}
\end{align}

\par
The special case, $\lambda=0$ (and $\alpha=0$ to have a finite $\xi$) is nothing but the Landau gauge in the conventional Lorentz gauge and the BRST and anti-BRST invariant operator $\mathcal{O}$ reduces to a simple form:
\begin{equation}
  \mathcal{O}' =(\Omega^{(D)})^{-1} \int d^D x \  \left[ 
 {1 \over 2} \mathscr{A}_\mu(x) \cdot \mathscr{A}^\mu{}(x)      \right] .
\end{equation}
Note that $\mathcal{O}'$ is invariant under the gauge transformation as well as the BRST and anti-BRST transformation.

\subsection{Renormalization of the composite operator}

Hereafter, we use the following notation to simplify the expressions.
\begin{equation}
  A :=\mathscr{A}^{\rm R} , \quad
  C :=\mathscr{C}^{\rm R}, \quad
 \bar{C} :=\bar{\mathscr{C}}^{\rm R} , \quad
 B :=\mathscr{B}^{\rm R} .
\end{equation}

\def\mathgraph#1{%
  \begin{array}{c}%
  \includegraphics[height=1cm]{#1}%
  \end{array}%
}
\def\mathgrapha#1{%
  \begin{array}{c}%
  \includegraphics[height=.7cm]{#1}%
  \end{array}%
}
\def\AAAA{%
  \begin{array}{c}
  \includegraphics[height=.2cm]{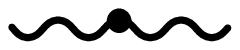}
  \end{array}
}
\def\CCCC{%
  \begin{array}{c}
  \includegraphics[height=.15cm]{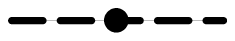}
  \end{array}
}
\def\AAAAi{\mathgraph{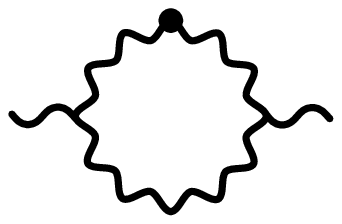}}
\def\AAAAii{\mathgraph{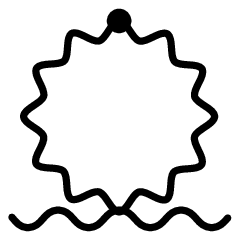}}
\def\AACCi{\mathgraph{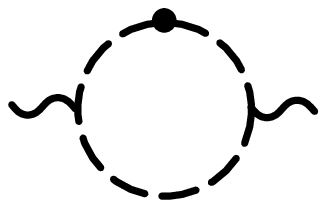}}
\def\AACCii{\mathgraph{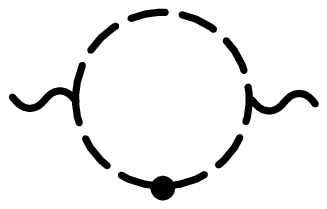}}
\def\CCAA{\mathgrapha{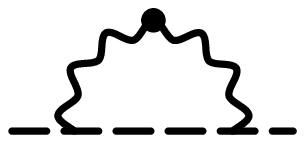}}
\def\CCCCi{\mathgrapha{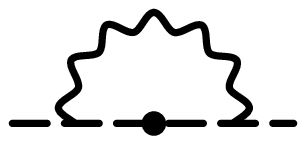}}
\def\CCCCii{\mathgraph{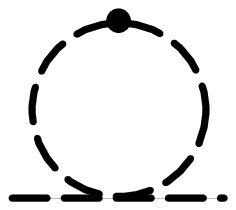}}

We consider the Green function of the fundamental fields with an insertion of the composite operator of mass dimension 2.
In the following, it is assumed that we have already finished the renormalization for the fundamental field in the perturbative theory.   
Therefore, we have only to consider the extra renormalization for the divergence coming from the inserted composite operators in the renormalized Green function.
In order to take into account the operator mixing among composite operators with the same mass dimension and the same quantum number, we must introduce the matrix of renormalization factors $Z_1, \cdots, Z_4$:
\begin{equation}
  \left(\begin{array}{c}
        \left[\frac12AA\right]_{\rm R} \\[1mm]
        [i\bar CC]_{\rm R}
        \end{array}
  \right)
 =\left(\begin{array}{cc}
        Z_1 & Z_2 \cr
        Z_3 & Z_4
        \end{array}
  \right) 
\left(\begin{array}{c}
      \left[\frac12AA\right] \\[1mm]
      [i\bar CC]
      \end{array}
\right) .
\label{rel}
\end{equation}
Then, to the lowest nontrivial order, we find
\begin{subequations}
\begin{alignat}{2}
\textstyle\left<AA\left[\frac12AA\right]\right>
 &=\AAAA &+\AAAAi &+\AAAAii
    +\ \cdots ,
    \\
\textstyle\left<AA\left[i\bar CC\right]\right>
 &=\quad\ \ 0 &+ \AACCi &+\AACCii
    +\ \cdots ,
    \\
\textstyle\left<i\bar CC\left[\frac12AA\right]\right>
 &=\quad\ \ 0 &+ \CCAA &%
    +\ \cdots ,
    \\
\textstyle\left<i\bar CC\left[i\bar CC\right]\right>
 &=\CCCC &+\CCCCi &+\CCCCii
    +\ \cdots ,
\end{alignat}
\end{subequations}
where 
we have used the Feynman rule:
\begin{subequations}
\begin{align}
 \AAAA =& \delta^{AB} ,
\\
 \CCCC =& i\delta^{AB} ,
\end{align}
\end{subequations}
with the dot denoting the insertion of a composite operator.

\par
We show that the divergences coming from the compositeness are absorbed by taking the four renormalization constants $Z_1$, $Z_2$, $Z_3$, $Z_4$ appropriately.
The first example is
\begin{align}
\textstyle\left<AA\left[\frac12AA\right]_{\rm R}\right>
 =\,& Z_1\textstyle\left<AA\left[\frac12AA\right]\right>
    +Z_2\textstyle\left<AA\left[i\bar CC\right]\right>
    \nonumber\\
 =\,& Z_1\left\{\AAAA+\AAAAi+\AAAAii
              +\cdots\right\}
    \nonumber\\
 &  +Z_2\left\{\AACCi+\AACCii
                    +\cdots\right\}
    \nonumber\\
 \equiv& 
    \AAAA .
\end{align}
Hence the lowest value of $Z_1$ is $1$:
\begin{equation}
Z_1=1+Z_1^{(1)}+\cdots.
\end{equation}
The second example is
\begin{align}
\textstyle\left<i\bar CC\left[\frac12AA\right]_{\rm R}\right>
 =\,& Z_1\textstyle\left<i\bar CC\left[\frac12AA\right]\right>
    +Z_2\textstyle\left<i\bar CC[i\bar CC]\right>
    \nonumber\\
 =\,& Z_1\left\{\CCAA
              +\cdots\right\}
    \nonumber\\
 & +Z_2\left\{\CCCC+\CCCCi+\CCCCii
               +\cdots\right\}
    \nonumber\\
 \equiv& 0 \ (\text{no divergence}) .
\end{align}
Hence $Z_2$ does not have the tree part and begins with the one-loop order:
\begin{equation}
Z_2=Z_2^{(1)}+\cdots.
\end{equation}
The third example is
\begin{align}
\textstyle\left<i\bar CC\left[i\bar CC\right]_{\rm R}\right>
 =\,& Z_3\textstyle\left<i\bar CC\left[\frac12AA\right]\right>
    +Z_4\textstyle\left<i\bar CC\left[i\bar CC\right]\right>
    \nonumber\\
 =\,& Z_3\left\{\CCAA
              +\cdots\right\}
    \nonumber\\
 &  +Z_4\left\{\CCCCi+\CCCCii
              +\cdots\right\}
    \nonumber\\
 \equiv &
    \CCCC .
\end{align}
Hence, $Z_4$ has the form:
\begin{equation}
Z_4=1+Z_4^{(1)}+\cdots.
\end{equation}
The fourth example is
\begin{align}
\textstyle\left<AA\left[i\bar CC\right]_{\rm R}\right>
 =\,&Z_3\textstyle\left<AA\left[\frac12AA\right]\right>
    +Z_4\textstyle\left<AA\left[i\bar CC\right]\right>
    \nonumber\\
 =\,&Z_3\left\{\AAAA+\AAAAi+\AAAAii
              +\cdots\right\}
    \nonumber\\
 & +Z_4\left\{\AACCi+\AACCii
                    +\cdots\right\}
    \nonumber\\
 \equiv& 0 \ (\text{no divergence}) .
\end{align}
Hence, $Z_3$ begins with the one-loop order:
\begin{equation}
Z_3=Z_3^{(1)}+\cdots.
\end{equation}
Therefore, up to one-loop order, the renormalization constants must satisfy the relationship:
\begin{subequations}
\begin{alignat}{3}
\begin{array}{c}
Z_1^{(1)} \\[-1mm] \AAAA
\end{array}
 &+\AAAAi &+&\AAAAii &=\,&0 ,
\label{eq:counter1}
\\
\begin{array}{c}
Z_2^{(1)} \\[-1mm] \CCCC
\end{array}
 &+\CCAA &&%
 &=\,&0 ,
\label{eq:counter2}
\\
\begin{array}{c}
Z_3^{(1)} \\[-1mm] \AAAA
\end{array}
 &+\AACCi &+&\AACCii &=\,&0 ,
\label{eq:counter3}
\\
\begin{array}{c}
Z_4^{(1)} \\[-1mm] \CCCC
\end{array}
 &+\CCCCi &+&\CCCCii &=\,&0 .
\label{eq:counter4}
\end{alignat}
\end{subequations}
The explicit calculations lead to the following divergent part.
\begin{align}
\AAAAi
 &\sim C_2(G) \delta^{AB}
      \left[3+\frac34\lambda(1+\lambda)\right]g_{\mu\nu}
      \frac{(g\mu^{-\epsilon})^2}{(4\pi)^2}
      \frac{1}\epsilon ,
\label{ins1}
\\
\AAAAii
 &\sim -3C_2(G) \delta^{AB}g_{\mu\nu}\frac{3+\lambda^2}4
      \frac{(g\mu^{-\epsilon})^2}{(4\pi)^2}
      \frac{1}\epsilon ,
\label{ins2}
\\
\CCAA
 &\sim iC_2(G) \delta^{AB}\xi(1-\xi)\lambda^2
    \frac{(g\mu^{-\epsilon})^2}{(4\pi)^2}\frac1\epsilon ,
\label{ins3}
\\
\AACCi
 &\sim -\frac14C_2(G) \delta^{AB}g_{\mu\nu}
     \frac{(g\mu^{-\epsilon})^2}{(4\pi)^2}\frac1\epsilon ,
\label{ins4}
\\
\CCCCi
 &\sim iC_2(G) \delta^{AB}\xi(1-\xi)\lambda
     \frac{(g\mu^{-\epsilon})^2}{(4\pi)^2}\frac1\epsilon ,
\label{ins5}
\\
\CCCCii
 &\sim -iC_2(G) \delta^{AB}\xi(1-\xi)\lambda
      \frac{(g\mu^{-\epsilon})^2}{(4\pi)^2}\frac1\epsilon .
\label{ins6}
\end{align}

Thus the renormalization constants for the composite operators are obtained as 
\begin{subequations}
\begin{align}
Z_1^{(1)}
 &=-\frac34(1+\lambda)
   C_2(G) \frac{(g\mu^{-\epsilon})^2}{(4\pi)^2}\frac1\epsilon ,
\\
Z_2^{(1)}
 &=-\lambda^2\xi(1-\xi)
   C_2(G) \frac{(g\mu^{-\epsilon})^2}{(4\pi)^2}\frac1\epsilon ,
\\
Z_3^{(1)}
 &=\frac12
   C_2(G) \frac{(g\mu^{-\epsilon})^2}{(4\pi)^2}\frac1\epsilon ,
\\
Z_4^{(1)}
 &=0 .
\end{align}
\end{subequations}

We pay attention to the renormalization constants of composite operators in light of the inverted relation of (\ref{rel}):
\begin{equation}
\left(\begin{array}{c}
      \left[\frac12AA\right] \\[1mm]
      [i\bar CC]
      \end{array}
\right)
 =\left(\begin{array}{cc}
        Z_1 & Z_2 \cr
        Z_3 & Z_4
        \end{array}
  \right)^{-1}
  \left(\begin{array}{c}
        \left[\frac12AA\right]_{\rm R} \\[1mm]
        [i\bar CC]_{\rm R}
        \end{array}
  \right)
 =\left(\begin{array}{cc}
        1-Z_1^{(1)} &  -Z_2^{(1)} \cr
         -Z_3^{(1)} & 1-Z_4^{(1)}
        \end{array}
  \right)
  \left(\begin{array}{c}
        \left[\frac12AA\right]_{\rm R} \\[1mm]
        [i\bar CC]_{\rm R}
        \end{array}
  \right) ,
\label{invr}
\end{equation}
This relation shows that there is an operator mixing between the gluon and ghost composite operators which are of mass dimension 2 and color singlet, as pointed out in \cite{Kondo01}.
In the absence of four-ghost interaction ($\xi=0$ or $\xi=1$), 
(\ref{ins3}), (\ref{ins5}) and (\ref{ins6}) vanish and hence
we have $Z_2^{(1)}=0=Z_4^{(1)}$.  In this case, there is no contribution from ghost for the renormalization of the gluon composite operator $\left[\frac12AA\right]$:
\begin{align}
  \left[\frac12AA\right] &=(1-Z_1^{(1)}) \left[\frac12AA\right]_{\rm R} ,
\\
 [i\bar CC] &=[i\bar CC]_{\rm R} -Z_3^{(1)} \left[\frac12AA\right]_{\rm R}  .
\end{align}
On the other hand, the ghost composite operator can not be finite without the mixing of the gluon composite operator.  
In the conventional Lorentz gauge fixing, therefore, we do not have to consider the contribution from ghost in treating the renormalization of the gluon composite operator $\left[\frac12AA\right]$ (at least in the one-loop level).

\subsection{Multiplicative renormalizability of the composite operator}
\par
Now we examine the multiplicative renormalizability of the composite operator $\mathcal{O}$.
Taking into account the renormalization of the fundamental field and the composite field (\ref{invr}), we obtain
\begin{align}
Q_0 :=\,& \frac12A_0A_0+ \lambda_0 i \bar C_0C_0
    \nonumber\\
 =\,& \left(1+Z_A^{(1)}\right)\frac12AA
    +\left(1+Z_\lambda^{(1)}\right)\left(1+Z_C^{(1)}\right) \lambda i \bar CC
    \nonumber\\
 =\,& \left(1+Z_A^{(1)}\right)
    \left\{\left(1-Z_1^{(1)}\right)\left[\textstyle\frac12AA\right]_{\rm R}
           -Z_2^{(1)}[i\bar CC]_{\rm R}\right\}
    \nonumber\\
 &   +\left(1+Z_\lambda^{(1)}\right)\left(1+Z_C^{(1)}\right)
 \lambda
     \left\{-Z_3^{(1)}\left[\textstyle\frac12AA\right]_{\rm R}
            +\left(1-Z_4^{(1)}\right)[i\bar CC]_{\rm R}\right\}
    \nonumber\\
 =\,& \left\{1+Z_A^{(1)}-Z_1^{(1)}-\lambda Z_3^{(1)}\right\}
    \left[\textstyle\frac12AA\right]_{\rm R}
    \nonumber\\
 &   + \left\{-\frac{Z_2^{(1)}}\lambda
                 +1+Z_\lambda^{(1)}+Z_C^{(1)}-Z_4^{(1)}\right\}
     \lambda [i\bar CC]_{\rm R} .
\end{align}
The multiplicative renormalizability holds (in the one-loop level) if and only if
\begin{equation}
  Z_Q^{(1)} :=Z_A^{(1)}-Z_1^{(1)}-\lambda Z_3^{(1)}
=-\frac{Z_2^{(1)}}\lambda
                  +Z_\lambda^{(1)}+Z_C^{(1)}-Z_4^{(1)} .
\label{cond}
\end{equation}
This is equivalent to the condition:
\begin{equation}
  \lambda  \left( \xi-{1 \over 2} \right)^2 =0 .
\end{equation}
If this condition is satisfied, the composite operator is multiplicatively renormalized as
\begin{align}
  Q_0 &= Z_Q \left( \left[\textstyle\frac12AA\right]_{\rm R}+ \lambda [i\bar CC]_{\rm R} \right) ,
\\
 Z_Q^{(1)} &= \left( {35 \over 12} - {1 \over 4} \lambda \right)
   C_2(G) \frac{(g\mu^{-\epsilon})^2}{(4\pi)^2}\frac1\epsilon .
\end{align}
In the case of $\lambda=0$, this result reduces to that of Boucaud et al. \cite{Boucaudetal01} without operator mixing.
\par
It should be remarked that the composite operator is not multiplicatively renormalizable, unless the renormalization of the composite operators $AA$ and $\bar{C}C$ are taken into account.
In fact, the multiplicative renormalizability of 
\begin{align}
Q_0 &:=\frac12A_0A_0+\lambda i \bar C_0C_0
    \nonumber\\
 &= \left(1+Z_A^{(1)}\right)\frac12AA
    +\left(1+Z_\lambda^{(1)}+Z_C^{(1)}\right) \lambda i \bar CC 
    + O(\hbar^2) ,
\end{align}
without the renormalization of the composite operator 
leads to the condition:
$
  Z_A^{(1)} - Z_\lambda^{(1)} - Z_C^{(1)} =0 ,
$
which reads
$
  \lambda \left[ \xi (\xi-1) + {1 \over 4} \right] ={3 \over 4} .
$
This curve does not have a definite meaning in the renormalization, since the curve is not along the RG flow.

\subsection{BRST invariance of the renormalized composite operator}

Finally, we show that the renormalized composite operator $\mathcal{O}^{\rm R}$ is invariant under the renormalized BRST and anti-BRST transformation.
By requiring that the renormalized BRST and anti-BRST transformations are nilpotent and anti-commute:
\begin{equation}
  \bm{\delta}_{\rm B}^{\rm R} \bm{\delta}_{\rm B}^{\rm R} \equiv 0 , \quad
\bar{\bm{\delta}}_{\rm B}^{\rm R} \bar{\bm{\delta}}_{\rm B}^{\rm R} \equiv 0 , \quad
 \bm{\delta}_{\rm B}^{\rm R} \bar{\bm{\delta}}_{\rm B}^{\rm R} + \bar{\bm{\delta}}_{\rm B}^{\rm R} \bm{\delta}_{\rm B}^{\rm R} \equiv 0 ,
\end{equation}
the renormalized BRST and anti-BRST transformation for the renormalized fields $A_\mu, C, \bar{C}, B$ are determined (by an    appropriate rescaling of $\mathscr{B}$ field) as \cite{BT82,Baulieu85}
\begin{subequations}
\begin{align}
 \bm{\delta}_{\rm B}^{\rm R} A_\mu(x) &=
X \mathscr{D}_\mu[A]^{\rm R}C(x) :=
 X[\partial_\mu C(x) + Z_A^{1/2} Z_g g_{\rm R} (A_\mu(x) \times C(x))] , \\
 \bm{\delta}_{\rm B}^{\rm R} C(x) &=-{1 \over 2}X Z_A^{1/2} Z_g g_{\rm R} (C(x) \times C(x)) ,
\\
 \bm{\delta}_{\rm B}^{\rm R} \bar{C}(x) &=i X B(x) ,
\\
 \bm{\delta}_{\rm B}^{\rm R} B(x) &=0 ,
\label{BRST3}
\end{align}
\end{subequations}
and  
\begin{subequations}
\begin{align}
 \bar{\bm{\delta}}_{\rm B}^{\rm R} A_\mu(x) &=
 \bar{X} \mathscr{D}_\mu[A]^{\rm R} \bar{C}(x) :=
\bar{X}[\partial_\mu \bar{C}(x) + Z_A^{1/2} Z_g g_{\rm R} (A_\mu(x) \times \bar{C}(x))] , \\
 \bar{\bm{\delta}}_{\rm B}^{\rm R} \bar{C}(x) &=-{1 \over 2}\bar{X} Z_A^{1/2} Z_g  g_{\rm R} (\bar{C}(x) \times \bar{C}(x)) ,
 \\
 \bar{\bm{\delta}}_{\rm B}^{\rm R} C(x) &=i \bar{X} \bar{B}(x) ,
\\
 \bar{\bm{\delta}}_{\rm B}^{\rm R} \bar{B}(x) &=0 ,
\label{BRST4}
\end{align}
\end{subequations}
where $X$ and $\bar{X}$ are arbitrary real numbers and
$\bar{B}$ is defined by
\begin{equation}
  \bar{B}(x) =-B(x) + iZ_A^{1/2} Z_g g_{\rm R} (C(x) \times \bar{C}(x)) .
\end{equation}
\par
The Lagrangian is written by making use of the renormalized BRST and anti-BRST transformation and the renormalized fields as
\begin{align}
  \mathscr{L}_{\rm YM}^{\rm tot} =\,& 
- {1 \over 4} Z_A (\partial_\mu A_\nu  - \partial_\nu A_\mu  + Z_g Z_A^{1/2} g_{\rm R} A_\mu \times A_\nu )^2
\nonumber\\&
+ {Z_C \over X\bar{X}}
 i\bm{\delta}_{\rm B}^{\rm R} \bar{\bm{\delta}}_{\rm B}^{\rm R}
 \left( {1 \over 2} A_\mu \cdot A^\mu -\frac{Z_CZ_\alpha}{Z_A} {\alpha_{\rm R} \over 2}i C \cdot \bar{C} \right) 
+ \frac{Z_C^2 Z_{\alpha'}}{Z_A}{\alpha_{\rm R}' \over 2} B \cdot B  .
\end{align}
This agrees with (\ref{rbLag}).
\par
We derive the condition for the renormalized composite operator ${\cal O}_{\rm R}$ to be invariant under the renormalized BRST transformation defined above.  We can write a finite composite operator of mass dimension 2 in the form (up to an overall constant):
\begin{equation}
   {\cal Q}_{\rm R} =\left[ \frac12A_\mu(x) \cdot A^\mu(x) \right]_{\rm R} + K_{\rm R} [i\bar{C}(x) \cdot C(x)]_{\rm R} ,
\end{equation}
where $K_{\rm R}$ is a finite function of the renormalized parameters, 
$g_{\rm R}, \xi_{\rm R}, \lambda$.
Performing the renormalized BRST transformation (\ref{BRST3}) 
after  the renormalization factors (\ref{invr}) of the composite operator are included, we obtain 
\begin{align}
   \bm{\delta}_{\rm B}^{\rm R}{\cal Q}_{\rm R} =\,& 
 \bm{\delta}_{\rm B}^{\rm R}  \Big\{ (Z_1 + K_{\rm R} Z_3)(\frac12 A_\mu \cdot A^\mu ) 
 + (Z_2 + K_{\rm R} Z_4)(i\bar{C} \cdot C ) \Big\} 
\nonumber\\
 =\,&  (Z_1 + K_{\rm R} Z_3) X \partial_\mu C \cdot A^\mu 
\nonumber\\
&+ (Z_2 +K_{\rm R} Z_4)\Big\{ i\bar{C} \cdot (X Z_A^{1/2} Z_g \frac{g}{2}C \times C ) 
 \nonumber\\
&+ X \left( \frac{Z_A}{Z_{C}Z_\lambda} \frac{1}{\lambda} \partial_\mu A^\mu  
-i Z_A^{1/2} Z_\xi \xi Z_g g C \times\bar{C} \right) \cdot C \Big\} .
\end{align}
For the right-hand-side to be a total derivative, we must require two conditions: 
1) the coefficient for the term 
$C\cdot (\bar{C} \times C)$ vanishes, 
2) the remaining terms containing the derivative are combined into a total derivative term.  The respective condition reads
\begin{align}
\frac{Z_A^{1/2} Z_g}{2}  =\,& Z_A^{1/2} Z_g  Z_\xi \xi  ,
\\
 Z_1 +K_{\rm R} Z_3   =\,& 
(Z_2 +K_{\rm R} Z_4) \frac{Z_A}{Z_{C}Z_\lambda} \frac{1}{\lambda} .
\end{align}
The first condition reduces to 
\begin{equation}
 \xi_0 =Z_\xi \xi =\frac12 .
\end{equation}
Since 
$Z_2, Z_3  \sim O(\hbar/\epsilon)$ and
$Z_1, Z_4  \sim 1+O(\hbar/\epsilon)$,
 the second condition yields for the $O(1)$ term: 
\begin{equation}
  K_{\rm R} =\lambda_{\rm R} ,
\end{equation}
and for the $O(1/\epsilon)$ term:
\begin{equation}
 Z^{(1)}_A - Z^{(1)}_1 - \lambda_{\rm R} Z^{(1)}_3  
 + \frac{Z^{(1)}_2}{\lambda_{\rm R}} - Z^{(1)}_\lambda - Z^{(1)}_C + Z^{(1)}_4 =0.
\label{ZallCondition}
\end{equation}
This condition is the same as (\ref{cond}).
In the Landau gauge $\alpha =\lambda =0 $, especially, the condition (\ref{ZallCondition})
reduces to 
$
  Z_2^{(1)} =0 .
$
This is automatically satisfied in this case.
\par


\section{Operator product expansion and vacuum condensate}
We apply the operator product expansion (OPE) or short distance expansion (SDE) to the gluon and ghost propagators.
The OPE is originally proposed as an operator relation by Wilson \cite{Wilson69}.  For example, the product of two scalar field operators  defined at different spacetime points is expanded as  
\begin{equation}
  \phi(x)\phi(y) \sim \sum_{i} F^{[\mathcal{O}_i]}(x-y) \left[\mathcal{O}_i \left( {x+y \over 2} \right) \right] ,  
\end{equation} 
where the composite operators $\{ \mathcal{O}_i \}$ form a complete set of renormalized local  operators.
The famous proof of OPE by Zimmermann \cite{Zimmermann70} was given in the framework of the 
perturbation theory. 
Quite recently, the OPE was rigorously proved as an operator relation by 
Bostelman\cite{Bostelman00}.%
\footnote{
The authors would like to thank Izumi Ojima for informing this reference.
}
According to the method \cite{Larsson85,TM83},  
the (Fourier transformed) Wilson coefficient $\tilde F^{[\phi_1 \cdots \phi_n]}(p)$ 
in the OPE: 
\begin{equation}
  \phi(x)\phi(y) \sim \sum_{n} F^{[\phi_1 \cdots \phi_n]}(x-y) 
\left[\phi_1 \cdots \phi_n \left( {x+y \over 2} \right) \right] ,
\end{equation}
can be calculated in perturbation theory by equating a $(2+n)$-point 
one-particle irreducible (1PI) Green's function --- where two of the external legs have hard 
momentum $p$ and the remaining $n$ external legs are assigned zero momentum $q=0$ --- with the 
Wilson coefficient times an $n$-point Green's function with an insertion of the relevant composite operator 
at zero momentum. 

\subsection{The OPE in the tree level}

\par
First, we consider the OPE of the inverse gluon propagator:
\begin{align}
  (D^{-1})^{AB}_{\mu\nu}(p) &=
    C^{[1]}{}_{\mu\nu}^{AB}(p) \langle 1 \rangle
    +  C^{[A^2]}{}_{\mu\nu}^{AB}(p)  \langle \frac12 A_\rho \cdot A^\rho \rangle
    + C^{[\bar{C} C]}{}_{\mu\nu}^{AB}(p)  \langle i\bar{C} \cdot C \rangle
    +\cdots ,
\label{invgluonOPE}
\end{align}
where the first Wilson coefficient is nothing but the bare inverse gluon propagator:
\begin{align}
 C^{[1]}{}_{\mu\nu}^{AB}(p) &=(D_0^{-1})_{\mu\nu}^{AB}(p) :=- p^2 (P^{\rm T}_{\mu\nu}+\lambda^{-1} 
P^{\rm L}_{\mu\nu} )\delta^{AB} \nonumber \\
  &=-p^2 \left( g_{\mu\nu}-\frac{p^\mu p^\nu}{p^2} +\lambda^{-1} \frac{p^\mu p^\nu}{p^2} 
\right) \delta^{AB} .
\end{align}
\par
The other Wilson coefficients are calculated in the perturbation theory from the diagrams:
\begin{align}
 i C^{[A^2]}_{\mu\nu} =&
  \begin{array}{c}
  \includegraphics[width=1.4cm]{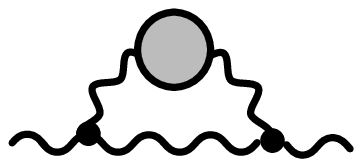}
  \end{array}
 + 
  \begin{array}{c}
  \includegraphics[width=1.0cm]{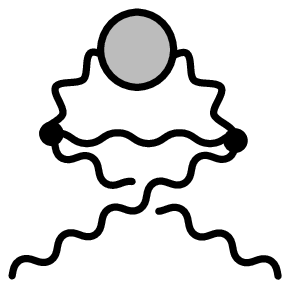}
  \end{array}
 + 
  \begin{array}{c}
  \includegraphics[width=1.0cm]{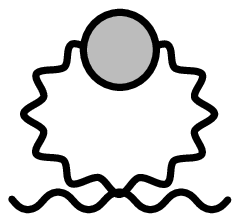}
  \end{array},
\\
 C^{[\bar{C} C]}_{\mu\nu} =&
  \begin{array}{c}
  \includegraphics[width=1.4cm]{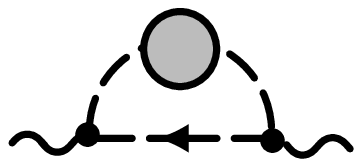}
  \end{array}
 + 
  \begin{array}{c}
  \includegraphics[width=1.4cm]{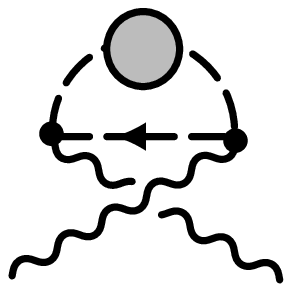}
  \end{array}
 .
\end{align}
In the diagram, two external legs have hard momentum $p$ and the ($n=2$) lines connected to a blob correspond to the external legs with zero momentum $q=0$.  
\par
The explicit calculation in the tree level yields the result (see Appendix for the details of 
calculations):
\begin{align}
 C^{[A^2]}{}_{\mu\nu}^{AB}(p) 
&= -\frac{N_c g^2}{2(N_c^2-1)} (1+\lambda) P^{\rm T}_{\mu\nu} \delta^{AB} ,
\\
  C^{[\bar{C}C]}{}_{\mu\nu}^{AB}(p) &= 2 \frac{N_c g^2}{(N_c^2-1)} \xi(1-\xi) P^{\rm L}_{\mu\nu} 
\delta^{AB} ,
\end{align}
where we have put $C_2(G)=N_c$ for simplicity.
Defining the vacuum polarization tensor of the gluon by
\begin{equation}
  (D^{-1})^{AB}_{\mu\nu}(p) 
 := (D_0^{-1})_{\mu\nu}^{AB}(p) +  \Pi^{AB}_{\mu\nu}(p) ,
\end{equation}
we obtain the vacuum polarization tensor of the gluon:
\begin{align}
  \Pi^{AB}_{\mu\nu}(p) =
   \frac{N_c g^2}{4(N_c^2-1)} \delta^{AB}
  \left\{  -(1+\lambda) P^{\rm T}_{\mu\nu} \langle A_\rho \cdot A^\rho \rangle
    + 2 D\xi(1-\xi) P^{\rm L}_{\mu\nu} \langle i\bar{C} \cdot C \rangle  
  \right\} .
\label{gluonVP}
\end{align}
It turns out that even the inclusion of the quartic ghost interaction does not affect the Wilson coefficient 
$C^{[A^2]}_{\mu\nu}$ at least in the tree level.  
For the Wilson coefficient $C^{[\bar{C}C]}_{\mu\nu}$, however, there is an extra contribution coming from the quartic ghost interaction, as 
suggested already in \cite{Kondo01}.
The non-zero Wilson coefficient 
$C^{[\bar{C}C]}$ due to the presence of the quartic ghost interaction ($\xi\not=0,1$) breaks 
the transversality of the gluon polarization tensor, i.e., $\Pi_{\mu\nu} \not=P^{\rm T}_{\mu\nu} \Pi$. 
This result does not contradict with the Slavnov-Taylor identity \cite{CF76,BT82,KIMS01}.
When $\xi =0$ (resp.~$\xi=1$), the ghost condensate 
$\langle i\bar{C} \cdot C \rangle$
can not appear in the OPE, 
since the gluon--ghost--anti-ghost  vertex 
(\ref{ggagv}) is proportional to the outgoing ghost (resp.~anti-ghost) momentum $p_\mu$ (resp.~$q_\mu$).
The above result (\ref{gluonVP}) suggests the existence of the effective gluon mass given by
\begin{equation}
  m_A^2 = -
   \frac{N_c g^2}{4(N_c^2-1)}  
   (1+\lambda)  \langle A_\rho \cdot A^\rho \rangle  .
   \label{gluonmass}
\end{equation}
Therefore, the gluon condensation of mass dimension 2 can be an origin of the gluon mass. 
The effect of higher orders will be investigated in the next subsection.  

\par
Next, we perform the OPE for the inverse ghost propagator:
\begin{align}
 -i (G^{-1})^{AB}(p) &=
    C^{[1]}_{AB}(p) \langle 1 \rangle
    + C^{[A^2]}_{AB}(p)  \langle \frac12 A_\rho \cdot A^\rho  \rangle
    + C^{[\bar{C} C]}_{AB}(p)  \langle i\bar{C} \cdot C \rangle
    +\cdots ,
\label{invghostOPE}
\end{align}
where the first Wilson coefficient agrees with the bare inverse ghost propagator:
\begin{align}
    C^{[1]}_{AB}(p) =-i (G_0^{-1})^{AB}(p) =-p^2\delta^{AB} .
\end{align}
The other Wilson coefficients are calculated from the diagrams,
\begin{align}
  - C^{[A^2]}_{\rm gh} =&
  \begin{array}{c}
  \includegraphics[width=1.8cm]{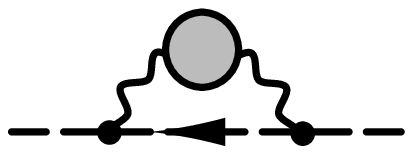}
  \end{array}
 + 
  \begin{array}{c}
  \includegraphics[width=1.2cm]{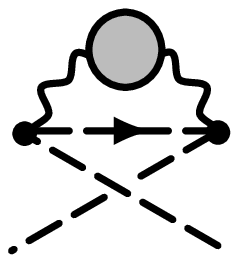}
  \end{array},
\\
  i C^{[\bar{C} C]}_g =&
  \begin{array}{c}
  \includegraphics[width=1.4cm]{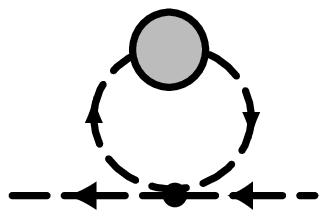}
  \end{array}
 +  
  \begin{array}{c}
  \includegraphics[width=1.8cm]{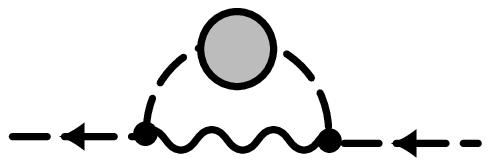}
  \end{array} ,
\end{align}
which yield the result:
\begin{align}
  C^{[A^2]}_{AB}(p) &=  \frac{N_c g^2}{2(N_c^2-1)}\delta^{AB},
\\
  C^{[\bar{C}C]}_{AB}(p) &=0 .
\end{align}
Here  the coefficient $C^{[\bar{C}C]}_{AB}$ vanishes due to cancellation, see Appendix.  
Defining the vacuum polarization tensor of the ghost by
\begin{equation}
  (G^{-1})^{AB}(p) 
 := (G_0^{-1})^{AB}(p) +   i \Pi_{\rm gh}^{AB}(p) ,
\end{equation}
the vacuum polarization for the ghost is obtained:
\begin{equation}
  \Pi_{\rm gh}^{AB}(p) =   \frac{N_c g^2}{4(N_c^2-1)} \delta^{AB}
    \langle A_\rho \cdot A^\rho \rangle  .
\label{ghostVP}
\end{equation}
We find that the ghost vacuum polarization has no contribution from the ghost-anti--ghost condensation even for $\xi\not=0,1$.  Thus we obtain the effective ghost mass:
\begin{equation}
  m_C^2 =\frac{N_c g^2}{4(N_c^2-1)}  
    \langle A_\rho \cdot A^\rho \rangle  .
 \label{ghostmass}
\end{equation}
This result shows that the gluon condensation of mass dimension 2 can also be an origin of the 
ghost mass.%
\footnote{In the Lorenz gauge, the effective gluon mass and ghost mass are generated by the 
gluon condensation of mass dimension 2 alone in the tree level.   
This is not the case if we include the high-order correction as will be shown in the next subsection.
In the MA gauge, on the contrary, two condensations from the off-diagonal gluon and off-diagonal ghost contribute to the effective off-diagonal gluon and ghost masses already in the tree level, see 
\cite{Kondo01,KS00}.
}
\par
The combination of gluon and ghost condensation appearing in the OPE 
is not BRST invariant in the sense explained in the previous section.  This is reasonable, 
since even the OPE of gauge invariant operators 
does not give a gauge invariant combination in the OPE, see e.g. \cite{LO92}.


\subsection{RG improvement of the OPE}

  One of the advantages of the OPE is that the momentum dependence of the Wilson coefficient 
is determined by the renormalization group (RG) equation.
  More accurately, the change of the Wilson coefficient under the RG transformation can be 
specified by the renormalization factors $Z$ 
which are to be calculated before the RG improvement of the OPE calculus.  Therefore, we can 
obtain the higher-order corrections for the momentum dependence of the coefficient without any 
explicit higher-order computations  (at least for the leading logarithmic  corrections).

\subsubsection{RG equation for Wilson coefficients}
\par
  We begin with an OPE relation in the momentum representation 
obtained by extracting composite operators up to mass dimension 2 
(we omit all the indices, since they are not essential in the following arguments): 
\begin{subequations}
\begin{align}
  -i \tilde A_{\rm R}(p) \tilde A_{\rm R}(-p) &= D_{\mbox{\tiny pert}}(p) [1] +
     F^A_1(p) \left[\frac{1}{2}A(0)A(0) \right]_{\rm R} + F^A_2(p) [i\bar C(0)C(0)]_{\rm R} +\cdots ,
\label{expa1}
\\
  \tilde{\bar{C}}_{\rm R}(p) \tilde C_{\rm R}(-p) &= -iG_{\mbox{\tiny pert}}(p) [1] +
     F^C_1(p) \left[\frac{1}{2}A(0)A(0) \right]_{\rm R} + F^C_2(p) [i\bar C(0)C(0)]_{\rm R}+\cdots ,
\label{expa2}
\end{align}
\end{subequations}
where $D_{\mbox{\tiny pert}}(p)$ and $G_{\mbox{\tiny pert}}$ denote respectively the perturbative gluon and ghost propagators in which the perturbative loop corrections are included besides the OPE contribution.  
\par
First, we try to rewrite all the field operators in both sides of (\ref{expa1}) and (\ref{expa2}) 
in terms of bare quantities. 
 Hereafter it is supposed that the Wilson coefficient and composite operators are defined based on the renormalization scheme depending on a certain parameter $\mu$ (corresponding to the mass scale), 
which is different from the BPHZ prescription at zero momentum, $q=0$.
In the actual calculations, we adopt the minimal subtraction (MS) scheme, although the resulting expressions can be translated into those of the momentum-space subtraction scheme (MOM).  
\par
 By making use of the $Z$ factors calculated already in the previous section, two OPE 
relations above are combined into a matrix form:
\begin{align}
   {\bf Z}_{\rm f}^{-1}
   \left(
    \begin{array}{c}
    -i\tilde A_0(p) \tilde A_0(-p) \cr
      \tilde{\bar{C}}_0(p) \tilde C_0(-p)
    \end{array}
   \right)
=  {\bf D}_{\mbox{\tiny pert}} 
 +   {\bf F} 
  \tilde{{\bf Z}}
    \left(
    \begin{array}{c}
    \frac{1}{2}A_0(0) A_0(0) \cr
    i \bar C_0(0) C_0(0)
    \end{array}
   \right) +\cdots ,
\end{align}
where we have defined the two by two matrices,
\begin{align}
 {\bf Z}_{\rm f} =
   \left(
    \begin{array}{cc}
    Z_A & 0 \cr
    0  & Z_C
    \end{array}
   \right) , 
\quad
 {\bf F} := \left(
    \begin{array}{cc}
    F^A_1(p) & F^A_2(p) \cr
    F^C_1(p)  & F^C_2(p)
    \end{array}
   \right)  ,
\quad
\tilde{{\bf Z}} =
   \left(
    \begin{array}{cc}
    Z_1 & Z_2 \cr
    Z_3  & Z_4
    \end{array}
   \right) 
{\bf Z}_{\rm f}^{-1} ,
\end{align}
and a column vector,
\begin{equation}
 {\bf D}_{\mbox{\tiny pert}} :=  \left(
     \begin{array}{c}
     D_{\mbox{\tiny pert}}(p) \cr
     -iG_{\mbox{\tiny pert}}(p) 
     \end{array}
    \right) .
\end{equation}
Introducing a matrix  ${\bf F}_0$ by
\footnote{Were it not for the renormalization of the composite operator,  ${\bf F}_0$ reduced to ${\bf F}$.}
\begin{align}
   {\bf F}_0 =
   {\bf Z}_{\rm f}
   {\bf F} 
   \tilde{{\bf Z}} 
:=
   \left(
    \begin{array}{cc}
    {F_0}^A_1(p) & {F_0}^A_2(p) \cr
    {F_0}^C_1(p)  & {F_0}^C_2(p)
    \end{array}
   \right) ,
\end{align}
we obtain an OPE relation among the bare quantities as
\begin{align}
   \left(
    \begin{array}{c}
   -i \tilde A_0(p) \tilde A_0(-p) \cr
      \tilde{\bar{C}}_0(p) \tilde C_0(-p)
    \end{array}
   \right)
 =  {\bf Z}_{\rm f} 
     {\bf D}_{\mbox{\tiny pert}}
  +   {\bf F}_0 
   \left(
    \begin{array}{c}
    \frac{1}{2}A_0(0) A_0(0) \cr
    i\bar C_0(0) C_0(0)
    \end{array}
   \right) +\cdots .
\label{relation}
\end{align}
\par
Second, we observe that the relation (\ref{relation})  should have no dependence on the renormalization point $\mu$.  
Hence, the first term on the right-hand-side of (\ref{relation}) is independent of $\mu$, i.e.,
\begin{equation}
     \mu\frac{d}{d \mu} ({\bf Z}_{\rm f} {\bf D}_{\mbox{\tiny pert}}) = 0 ,  
\label{indep1}
\end{equation}
and the coefficient ${\bf F}_0$ in the second term is also independent of $\mu$, i.e,
\begin{equation}
  \mu\frac{d}{d \mu} {\bf F}_0 
  = \mu\frac{d}{d \mu} ({\bf Z}_{\rm f} {\bf F} \tilde{{\bf Z}}) = 0 . 
\label{indep2}
\end{equation}
We multiply (\ref{indep2}) by ${\bf Z}_{\rm f}^{-1}$ from the left and by $\tilde{{\bf Z}}^{-1} $ 
from the right to obtain 
\begin{align}
  \left[ 
     \mu \frac{\partial}{\partial \mu} 
     +\sum_i \beta_i(\alpha) \frac{\partial}{\partial \alpha_i} 
 \right] {\bf F}
 +  {\bf Z}_{\rm f}^{-1} \left(\mu \frac{d}{d\mu}{\bf Z}_{\rm f} \right) {\bf F}
 + {\bf F} \left(\mu \frac{d}{d\mu} \tilde{{\bf Z}} \right) \tilde{{\bf Z}}^{-1} =0 \ ,
\end{align}
where $\alpha_i$ denotes the parameters of the theory $(g_{\rm R}, \xi_{\rm R}, \lambda_{\rm R})$, 
and $\beta_i$ denotes the corresponding RG function
$\beta_i(\alpha) := \mu\frac{\partial}{\partial \mu} \alpha_i$. 
Here we have used a fact that 
$ 
\mu \frac{\partial}{\partial \mu}+\sum_i \beta_i({\alpha(\mu)}) \frac{\partial}{\partial 
\alpha_i} 
$ 
is just the ordinary differential operator $\mu\frac{d}{d\mu}$.

\par
  Defining the RG function (matrix) $\bm{\gamma}_{\rm f}, \tilde{\bm{\gamma}}$  from ${\bf Z}_{\rm f},\ 
\tilde{{\bf Z}}$ by
\begin{align}
  \mu \frac{d}{d \mu} {\bf Z}_{\rm f} := {\bf Z}_{\rm f}  \bm{\gamma}_{\rm f} , 
\quad 
  \mu \frac{d}{d \mu} \tilde{{\bf Z}} := \tilde{\bm{\gamma}} \tilde{{\bf Z}}  ,
\label{gam}
\end{align}
we obtain the RG equation for the matrix ${\bf F}$ of the Wilson coefficients:
\begin{align}
  \Bigg[ 
     \mu \frac{\partial}{\partial \mu} 
     +\sum_i \beta_i(\alpha) \frac{\partial}{\partial \alpha_i} 
 \Bigg] {\bf F}(p, \alpha, \mu)
 +  \bm{\gamma}_{\rm f} {\bf F}(p, \alpha, \mu)
 + {\bf F}(p, \alpha, \mu) \tilde{\bm{\gamma}} =0 \ .
\label{OPERGeq}
\end{align}
\par
Similarly, we can show that ${\bf D}_{\mbox{\tiny pert}}$ obeys the RG equation:
\begin{align}
   \Bigg[
      \mu \frac{\partial}{\partial \mu}
      +\sum_i \beta_i({\alpha}) \frac{\partial}{\partial \alpha_i}
  \Bigg] {\bf D}_{\mbox{\tiny pert}}(p, \alpha, \mu)
  +  \bm{\gamma}_{\rm f} {\bf D}_{\mbox{\tiny pert}}(p, \alpha, \mu)
  =0 .
\label{OPERGv}
\end{align}

\subsubsection{Solving the RG equation}

\par
Now we proceed to solve the RG equation just obtained.
A simple dimensional analysis leads to a relation, 
${\bf F}(\kappa p, \alpha, \kappa \mu) 
= \kappa^{d_{\rm f}} {\bf F}\left(p, \alpha,  \mu \right)
$
 which is equivalent to the relation:
\begin{equation}
  {\bf F}(\kappa p, \alpha, \mu) = \kappa^{d_{\rm f}} {\bf F}\left(p, \alpha, {\mu \over \kappa} \right) ,
\end{equation}
where $d_{\rm f}$ is the canonical dimension of ${\bf F}$.  Hence, ${\bf F}$ satisfies
\begin{align}
  \left[ 
  \kappa \frac{\partial}{\partial \kappa}
     + \mu \frac{\partial}{\partial \mu}   - d_{F}
 \right] {\bf F}(\kappa p, \alpha, \mu) = 0 .
\label{OPERGeq2}
\end{align}
We use this equation to eliminate $\mu \frac{\partial}{\partial \mu}$ in (\ref{OPERGeq}) to obtain
\begin{align}
  \left[ 
    \kappa \frac{\partial}{\partial \kappa}
     -\sum_i \beta_i({\alpha}) \frac{\partial}{\partial \alpha_i}     - d_{F}
 \right] {\bf F}(\kappa p, \alpha, \mu)
 -  \bm{\gamma}_{\rm f} {\bf F}(\kappa p, \alpha, \mu)
 - {\bf F}(\kappa p, \alpha, \mu) \tilde{\bm{\gamma}} =0 \ .
\label{OPERGeq3}
\end{align}
This is the homogeneous RG equation of Weinberg-'t Hooft type \cite{homoRG} which is adequate for the mass-independent renormalization method.
\par
By the standard method \cite{Weinberg96,Kugo89}, the general solution of the RG equation (\ref{OPERGeq3}) is given by 
\begin{align}
  {\bf F}\left(\kappa p, \alpha, \mu \right) 
 = \kappa^{-4}
     \exp \left\{ \int_{1}^\kappa d\kappa' \frac{\bm{\gamma}_{\rm f}(\kappa')}{\kappa'} \right\} 
     {\bf F}(p,\bar{\alpha}(\kappa),\mu)
    \exp \left\{ \int_{1}^\kappa d\kappa' \frac{\tilde{\bm{\gamma}}(\kappa')}{\kappa'} \right\}  ,
 \label{OPERGsoln}  
\end{align}
where we have imposed the boundary condition:
$\bar{\alpha}(\kappa=1)=\alpha$.

 The similar consideration yields the general solution of the RG equation (\ref{OPERGv}):
\begin{align}
   {\bf D}_{\mbox{\tiny pert}}\left(\kappa p, \alpha, \mu \right)
  = \kappa^{-2}
     \exp \left\{ \int_{1}^\kappa d\kappa' \frac{\bm{\gamma}_{\rm f}(\kappa')}{\kappa'} \right\} 
     {\bf D}_{\mbox{\tiny pert}}\left(p, \alpha, \mu \right) .
\label{OPERGD}
\end{align}
\par
Once we know the $Z$ factors of the fundamental field and of the composite operator, it is 
easy to calculate $\bm{\gamma}_{\rm f},\  \tilde{\bm{\gamma}}$ according to (\ref{gam}). 
If the integrations in the arguments of the exponential in 
(\ref{OPERGsoln}) and (\ref{OPERGD}) are performed, 
the $\kappa$ dependence of the solution will be exactly determined. 
However,  $Z$ factors are obtained in terms of renormalized parameters $g_{\rm R}, \xi_{\rm R}, \lambda_{\rm R}$ 
and hence depend implicitly on $\kappa$ through them. 
This fact makes the analysis more difficult in general.

\subsubsection{Solution around the UV fixed point B}

\par
 We can calculate $\bm{\gamma}_{\rm f},\  \tilde{\bm{\gamma}}$ up to $O(\hbar)$, since we have known all the $Z$ factors of the fundamental field and of the composite operator up to $O(\hbar)$ in this paper. 
In the high-energy limit $\kappa \rightarrow \infty$, it is expected that the 
solution can be explicitly obtained in the neighborhood of the non-trivial UV stable fixed 
point in the parameter space, due to asymptotic freedom of the Yang-Mills theory, i.e., 
$\bar{g}(\kappa) \rightarrow \bar{g}_{\infty}=0$ as $\kappa \rightarrow \infty$. 
\par
In the three-dimensional parameter space $g_{\rm R},  \xi_{\rm R}, \lambda_{\rm R}$, actually, we have found 
that all the points are flowing into the UV fixed point B in the UV limit except for some 
lines that have higher symmetry.  
 On the other hand, within the perturbation theory using the dimensional regularization, 
the $\mu$ dependent loop correction of all the $Z$ factors always 
appears with a factor of $O(g_{\rm R}^2)$. 
Therefore, the RG function $\gamma$ as an element of the matrix $\bm{\gamma}$ defined by 
differentiating the $Z$ factor with respect to $\mu$ is accompanied by $g_{\rm R}^2$ to the 
$O(\hbar)$, 
like $\gamma \sim g_{\rm R}^2 f(\xi,\lambda)\hbar +O(\hbar^2)$. 
 If the polynomial function $f(\xi,\lambda)$ in the above expression 
has a nonvanishing value at the fixed point  $(\xi^*, \lambda^*)$ , 
the $\mu$ dependence of $\gamma=g^2 f$ is governed by $g^2$ alone
and hence we can replace $f(\xi,\lambda)$ with the constant $f(\xi^*, \lambda^*)$ at the UV 
fixed point. 
 By substituting the fixed-point values 
$ \lambda_{\rm R}^{*}=26/3,\  \xi^{*}_{\rm R}=1/2 $  into  $\xi,\  \lambda$,  the  $Z$ factors become
\begin{align}
 &  Z^*_A = 1 -\frac{13}{6} \frac{g^2 N_c}{16 \pi^2} \frac{\mu^{-2\epsilon}}{\epsilon}, \ 
 &  Z^*_C &= 1 -\frac{17}{12} \frac{g^2 N_c}{16 \pi^2} \frac{\mu^{-2\epsilon}}{\epsilon}, \  
\nonumber\\
 &  Z^*_1 = 1 -\frac{29}{4} \frac{g^2 N_c}{16 \pi^2} \frac{\mu^{-2\epsilon}}{\epsilon}, \ 
 &  Z^*_2 &= -\frac14\left(\frac{26}{3}\right)^2 \frac{g^2 N_c}{16 \pi^2} 
\frac{\mu^{-2\epsilon}}{\epsilon}, \ \nonumber\\
 &  Z^*_3 = \frac{1}{2} \frac{g^2 N_c}{16 \pi^2} \frac{\mu^{-2\epsilon}}{\epsilon}, \ 
 &  Z^*_4 &= 1 ,
\end{align}
which yield the matrix of the renormalization group function:
\begin{align}
     \bm{\gamma}^*_{\rm f}(g) = \frac{g^2 N_c}{8 \pi^2}
   \begin{pmatrix}
     \frac{13}{6} & 0\\
     0 & \frac{17}{12}
   \end{pmatrix}  
  ,\ \ \ 
\tilde{\bm{\gamma}}^*(g) = \frac{g^2 N_c}{8 \pi^2}
   \begin{pmatrix}
     \frac{61}{12}  & \frac14 \big(\frac{26}{3}\big)^2\\[2mm]
     -\frac12 & -\frac{17}{12}
   \end{pmatrix} .
 \label{gammafix}
\end{align}

\par
Furthermore, we define the coefficient matrix   
${\bf C}_{\gamma_{\rm f}}$ and ${\bf C}_{\tilde \gamma}$ in  (\ref{gammafix}) by
\begin{align}
  \bm{\gamma}^*_{\rm f}(g) := g^2 {\bf C}_{\gamma_{\rm f}},\ \ \ \  
  \tilde{\bm{\gamma}}^*(g) := g^2 {\bf 
C}_{\tilde \gamma} \ .
\end{align}
 By taking into account the RG equation $\mu \frac{d}{d\mu} g = 
-\frac{b}{8\pi^2}g^3 $ \ 
$(b = \frac{11}{6}N_c)$ and the resulting relation:
$
   \frac{d}{d\mu} \ln g^2 
     = \frac{2}{g} \frac{d}{d\mu} g 
     = -\frac{2b}{8\pi^2}\frac{g^2}{\mu} , 
$
the nontrivial integration of (\ref{OPERGsoln}) 
can be performed as 
\begin{align}
   \int^{\kappa}_{1} d\kappa' \frac{\bm{\gamma}(\bar{g}(\kappa'))}{\kappa'} 
     = \int^{\kappa}_{1} d\kappa' {\bf C}_{\gamma} \frac{(\bar{g}(\kappa'))^2}{\kappa'}
     =  {\bf C}_{\gamma} \frac{8\pi^2}{2b} 
 \ln {\bar{g}^2(1) \over \bar{g}^2(\kappa)} .
 \label{gammasoln} 
\end{align}
Hence the solution becomes  
\begin{equation}
  {\bf F}(\kappa p,\alpha,\mu) = \kappa^{-4} 
\left( {\bar{g}^2(1) \over \bar{g}^2(\kappa)}\right)^{{\bf C}_{\gamma_{\rm f}}\frac{8\pi^2}{2b}}  
({\bf F}(p,\bar{\alpha}(\kappa),\mu))
\left({\bar{g}^2(1) \over \bar{g}^2(\kappa)}\right)^{{\bf C}_{\tilde \gamma}\frac{8\pi^2}{2b}}  .
\label{solution}
\end{equation}
The $\kappa$ dependence of $\bar{g}^2$ is obtained by solving its 
RG equation as 
$\bar{g}^2(\kappa) \sim [\frac{2b}{8\pi^2} \ln \kappa]^{-1}$
for large $\kappa$. 
Substituting (\ref{gammasoln}) into (\ref{OPERGsoln}), therefore,
we determine the $\ln \kappa$ dependence of the solution for large $\kappa$:
\begin{equation}
  {\bf F}(\kappa p,\alpha,\mu) = \kappa^{-4} 
      \left( \ln \kappa \right)^{{\bf C}_{\gamma_{\rm f}}\frac{8\pi^2}{2b}}
        ({\bf F}(p,\bar{\alpha}(\kappa),\mu))
      \left( \ln \kappa \right)^{{\bf C}_{\tilde \gamma}\frac{8\pi^2}{2b}} .
\end{equation}

\par
  In order to cast the matrix power of $\ln \kappa$ into a more tractable form, 
we shall diagonalize the matrix ${\bf C}_{\tilde \gamma}$ in such a way that ${\bf S}$ 
diagonalizes ${\bf C}_{\tilde \gamma}$ by the similarity transformation ${\bf C}_{\tilde 
\gamma} \rightarrow {\bf S}^{-1} {\bf C}_{\tilde \gamma} {\bf S}$.
Such a matrix ${\bf S}$ and the diagonalized matrix are given by
\begin{align}
  {\bf S} = 
   \begin{pmatrix}
     -\frac{13}{3} & -\frac{26}{3}\\
     1 & 1
   \end{pmatrix}
,\ \ \ 
  {\bf S}^{-1} {\bf C}_{\tilde \gamma} {\bf S} =
   \frac{N_c}{8\pi^2}
   \begin{pmatrix}
     \frac{3}{4} & 0\\
     0 & \frac{35}{12}
   \end{pmatrix} .
\end{align}

This diagonalization corresponds to redefining the combination between two composite operators 
of mass dimension 2, i.e., $\frac{1}{2}A(0)A(0)$ and $i\bar C(0)C(0)$, by multiplying ${\bf S}^{-1}$:
\begin{align}
  \begin{pmatrix}
    {\cal Q}_1 \\
    {\cal Q}_2
  \end{pmatrix}
 = 
  S^{-1}
  \begin{pmatrix}
    \frac12 A^2 \\
    i\bar C C
  \end{pmatrix}
 =
  \frac{3}{13}
  \begin{pmatrix}
    \frac12 A^2 +\frac{26}{3}i\bar C C \\
    -\frac12 A^2 -\frac{13}{3}i\bar C C
  \end{pmatrix} .
\end{align}

 Inserting the identity matrix ${\bf 1}={\bf S} {\bf S}^{-1}$ appropriately, the solution (\ref{solution}) is 
rewritten as 
\begin{align}
   {\bf F}(\kappa p,\alpha,\mu)  = \kappa^{-4} 
        \left({\bar{g}^2(1) \over \bar{g}^2(\kappa)}\right)^{{\bf C}_{\gamma_{\rm f}}\frac{8\pi^2}{2b}}
           {\bf F}(p,\bar{\alpha},\mu)   {\bf S}
       {\bf S}^{-1} \left({\bar{g}^2(1) \over \bar{g}^2(\kappa)}\right)^{{\bf 
C}_{\tilde 
\gamma}\frac{8\pi^2}{2b}} {\bf S} {\bf S}^{-1}.
\end{align}
Now both ${\bf C}_{\gamma_{f}}$ and  ${\bf S}^{-1} {\bf C}_{\tilde \gamma} {\bf S}$ are 
diagonal.  
Hence we can write down the power explicitly as
\begin{align}
   {\bf F}(\kappa p)  &= \kappa^{-4}
      \begin{pmatrix}
         \left({\bar{g}^2(1) \over \bar{g}^2(\kappa)}\right)^{\frac{13}{6}\frac{N_c}{2b}}   & 
0\\
         0 & \left({\bar{g}^2(1) \over \bar{g}^2(\kappa)}\right)^{\frac{17}{12}\frac{N_c}{2b}} \\
      \end{pmatrix}
       {\bf T}(p)   {\bf S}
      \begin{pmatrix}
         \left({\bar{g}^2(1) \over \bar{g}^2(\kappa)}\right)^{\frac{3}{4}\frac{N_c}{2b}} & 0 \\
         0 & \left({\bar{g}^2(1) \over \bar{g}^2(\kappa)}\right)^{\frac{35}{12}\frac{N_c}{2b}} \\
      \end{pmatrix}
    {\bf S}^{-1} .
\end{align}
Here we impose a condition that ${\bf T}(p):={\bf F}(p,\bar{\alpha}(\kappa),\mu)$ coincides with the Wilson coefficient in the tree level  
 obtained in the previous section in which the coupling constant is replaced with the running coupling constant $\bar{\alpha}(\kappa)$. 
Note that ${\bf F}$ is the Wilson coefficient of the Green function (not of the one-particle irreducible (1PI) function).%
\footnote{
Except for the Landau gauge in which any operator mixing does not occur, 
a linear combination of different powers of $\ln \kappa$ appears in the solution,
and its combination coefficients cannot be completely determined by perturbation theory alone. 
But it is important to note that a fitting of the analytical result with the simulation data 
(or experimental data) can determine 
the asymptotic behavior of ${\bf F}$  completely as discussed in the next subsection.
}
Hence we put
\begin{align}
  {\bf T}(p) &=
    \begin{pmatrix}
       T_1(p) & T_2(p) \\
       T_3(p) & T_4(p) \\
    \end{pmatrix} 
\nonumber\\
  &=
     \frac{N_c \bar{g}^2(\kappa)}{2(N_c^2-1)}
    \begin{pmatrix}
       -(iD_0)^2 (1+\lambda)P_{\rm T} & (iD_0)^2 4\xi(1-\xi)P_{\rm L} \\
       (iG_0)^2  & 0 \\
    \end{pmatrix} \ \ .
\end{align}
We notice that  each element $T_1, \cdots, T_4$ of 
${\bf T}(p)$ 
 brings an extra $\ln \kappa$ factor to ${\bf F}$ through $\bar{g}^2(\kappa) \sim \frac{1}{\ln \kappa}$.
Therefore, the OPE correction up to dimension 2 operators reads 
\begin{align}
   {\bf F}(p)  
  \begin{pmatrix}
    \frac12 A^2 \\
    i\bar C C
  \end{pmatrix} 
 = &  
      \begin{pmatrix}
         (-\frac{13}{3}T_1 +T_2) 
           (\frac{\ln p/\Lambda_{\rm QCD}}{\ln \mu/\Lambda_{\rm QCD}})^{\frac{35}{12}\frac{N_c}{2b}}
       & (-\frac{26}{3}T_1 +T_2) 
           (\frac{\ln p/\Lambda_{\rm QCD}}{\ln \mu/\Lambda_{\rm QCD}})^{\frac{61}{12}\frac{N_c}{2b}}\\
         -\frac{13}{3}T_3  
           (\frac{\ln p/\Lambda_{\rm QCD}}{\ln \mu/\Lambda_{\rm QCD}})^{\frac{13}{6}\frac{N_c}{2b}}
       & -\frac{26}{3}T_3 
           (\frac{\ln p/\Lambda_{\rm QCD}}{\ln \mu/\Lambda_{\rm QCD}})^{\frac{13}{3}\frac{N_c}{2b}}
      \end{pmatrix}
  \begin{pmatrix}
    {\cal Q}_1 \\
    {\cal Q}_2 
  \end{pmatrix}  ,
\label{recomb}
\end{align}
where we have used $T_4=0$.  
Here we have used the translation rule from the MS scheme to the MOM scheme:
\begin{equation}
  {\bar{g}^2(1) \over \bar{g}^2(\kappa)} \longrightarrow 
  \frac{\ln p/\Lambda_{\rm QCD}}{\ln \mu/\Lambda_{\rm QCD}} .
\label{rule}
\end{equation}

\par
Among the terms with various powers of $\ln \kappa$, 
the largest-power term (corresponding to the largest eigenvalue of the matrix ${\bf C}_\gamma$) is dominant in the UV limit 
$(\kappa \gg 1)$.
Extracting this $\ln \kappa$ contribution, we can simplify the Wilson coefficient of 1PI function 
in the UV limit as
\begin{align}
  {\bf C}^{\rm 1PI} & = 
   \begin{pmatrix}
     C^{[A^2]}_{\mbox{\tiny gl}} & C^{[\bar CC]}_{\mbox{\tiny gl}} \\
     C^{[A^2]}_{\mbox{\tiny gh}} & C^{[\bar CC]}_{\mbox{\tiny gh}}
   \end{pmatrix}
  = 
   \begin{pmatrix}
     (iD_{\mbox{\tiny pert}})^{-2} & 0 \\
     0 & (iG_{\mbox{\tiny pert}})^{-2}
   \end{pmatrix} {\bf F}
\\
   & = 
      \frac{8\pi^2}{2b}\frac{N_c}{(N_c^2-1)}  
   \begin{pmatrix}
     (D_{\mbox{\tiny pert}}/D_0)^{-2} & 0 \\
     0 & (G_{\mbox{\tiny pert}}/G_0)^{-2}
   \end{pmatrix}
     \nonumber \\
      \times &
      \begin{pmatrix}
        \frac{13(-1-\lambda)P_{\rm T}  - 6\xi(1-\xi)P_{\rm L}}{13}
          \frac{ {(\ln \frac{p}{\Lambda_{\rm QCD}})}^{\frac{61}{12}\frac{N_c}{2b} -1} }
               { {(\ln \frac{\mu}{\Lambda_{\rm QCD}})}^{\frac{61}{12}\frac{N_c}{2b}} }  &
        \frac{13(-1-\lambda)P_{\rm T}  - 6\xi(1-\xi)P_{\rm L}}{3}
          \frac{ {(\ln \frac{p}{\Lambda_{\rm QCD}})}^{\frac{61}{12}\frac{N_c}{2b} -1} }
               { {(\ln \frac{\mu}{\Lambda_{\rm QCD}})}^{\frac{61}{12}\frac{N_c}{2b}} }  \\
        -1
          \frac{ {(\ln \frac{p}{\Lambda_{\rm QCD}})}^{\frac{26}{6}\frac{N_c}{2b} -1} }
               { {(\ln \frac{\mu}{\Lambda_{\rm QCD}})}^{\frac{26}{6}\frac{N_c}{2b}} }  &
        -\frac{13}{3}
          \frac{ {(\ln \frac{p}{\Lambda_{\rm QCD}})}^{\frac{26}{6}\frac{N_c}{2b} -1} }
               { {(\ln \frac{\mu}{\Lambda_{\rm QCD}})}^{\frac{26}{6}\frac{N_c}{2b}} }  
      \end{pmatrix} .
\label{impC}
\end{align}

In the similar way, we obtain
\begin{align}
    {\bf D}_{\mbox{\tiny pert}}(\kappa p,\alpha,\mu)  &= \kappa^{-2}
       \begin{pmatrix}
          \left({\bar{g}^2(1) \over \bar{g}^2(\kappa)}\right)^{\frac{13}{6}\frac{N_c}{2b}}  &
0\\
          0 & \left({\bar{g}^2(1) \over \bar{g}^2(\kappa)}\right)^{\frac{17}{12}\frac{N_c}{2b}} \\
       \end{pmatrix}
        {\bf D}_{t}(p) ,
\end{align}
where the tree expression is given by
\begin{align}
   {\bf D}_{t}(p)
   =
     \begin{pmatrix}
        D_0(p) \\
        -iG_0(p) \\
     \end{pmatrix}
   = 
     \begin{pmatrix}
        -\frac{1}{p^2}(P_{\rm T} +\lambda P_{\rm L})  \\
         \frac{1}{p^2} 
     \end{pmatrix}  .
\end{align}

\subsubsection{The solution at the conventional Landau gauge }

\par
Finally, we consider the OPE on the line A of the fixed points (corresponding to the 
conventional Landau gauge), 
the RG matrices read
\begin{align}
      \bm{\gamma}^*_{\rm f} = g^2 C_{\gamma_{\rm f}} 
= \frac{g^2 N_c}{8 \pi^2}
    \begin{pmatrix}
      -\frac{13}{6} & 0\\
      0 & -\frac{3}{4}
    \end{pmatrix}
   ,\ \ \
\tilde{\bm{\gamma}}^* =  g^2 C_{\tilde \gamma} 
= \frac{g^2 N_c}{8 \pi^2}
    \begin{pmatrix}
      \frac{35}{12}  & 0\\
      -\frac12 & \frac{3}{4}
    \end{pmatrix} .
  \label{gammafix2}
\end{align}
The diagonalization can be performed as
\begin{align}
   {\bf S} =
    \begin{pmatrix}
      0 & -\frac{13}{3}\\
      1 & 1
    \end{pmatrix}
,\ \ \
   {\bf S}^{-1} {\bf C}_{\tilde \gamma} {\bf S} =
    \frac{N_c}{8\pi^2}
    \begin{pmatrix}
      \frac{3}{4} & 0\\
      0 & \frac{35}{12}
    \end{pmatrix} .
\end{align}
The eigenvalues of ${\bf C}_{\tilde \gamma}$ are the same as those at the fixed point B.
Therefore, we obtain the Wilson coefficient $C^{[A^2]}_{\mu\nu}$
between $\langle A_\mu(p) A_\nu(-p)\rangle^{-1}$ and $\langle (A(0))^2\rangle$
and  $C^{[\bar CC]}$
between $\langle C(p) \bar{C}(-p)\rangle^{-1}$ and $\langle (A(0))^2\rangle$: 
\begin{align}
   {\bf F}(\kappa p) = \kappa^{-4}
     \begin{pmatrix}
      T_1(p) \left( {\bar{g}^2(1) \over \bar{g}^2(\kappa)}\right)^{\frac{3}{4}\frac{N_c}{2b}} & 0\\
      T_3(p) \left( {\bar{g}^2(1) \over \bar{g}^2(\kappa)}\right)^{\frac{13}{6}\frac{N_c}{2b}} & 0
     \end{pmatrix} ,
\end{align}
where no mixing between gluon and ghost occurs due to $T_2=0$ in addition to $T_4=0$.  
The coefficient of the 1PI OPE read
\begin{align}
  \begin{pmatrix}
   C^{[A^2]}_{\mbox{\tiny gl}} & C^{[\bar CC]}_{\mbox{\tiny gl}} \\
   C^{[A^2]}_{\mbox{\tiny gh}} & C^{[\bar CC]}_{\mbox{\tiny gh}} 
  \end{pmatrix}
 = 
  \frac{8\pi^2}{2b}\frac{N_c}{2(N_c^2-1)}
  \begin{pmatrix}
     - (\frac{D_{\mbox{\tiny pert}}}{D_0})^{-2} 
          \frac{ {(\ln \frac{p}{\Lambda_{\rm QCD}})}^{\frac{3}{4}\frac{N_c}{2b} -1} }
               { {(\ln \frac{\mu}{\Lambda_{\rm QCD}})}^{\frac{3}{4}\frac{N_c}{2b}} }  & 0\\
     (\frac{G_{\mbox{\tiny pert}}}{G_0})^{-2} 
          \frac{ {(\ln \frac{p}{\Lambda_{\rm QCD}})}^{\frac{13}{6}\frac{N_c}{2b} -1} }
               { {(\ln \frac{\mu}{\Lambda_{\rm QCD}})}^{\frac{13}{6}\frac{N_c}{2b}} }  & 0
  \end{pmatrix} ,
\end{align}
where
\begin{align}
    {\bf D}_{\mbox{\tiny pert}}(p)  &= 
       \begin{pmatrix}
           \left( \frac{\ln p/\Lambda_{\rm QCD}}{\ln \mu/\Lambda_{\rm QCD}}\right)^{-\frac{13}{6}\frac{N_c}{2b}}   &
0\\
          0 & \left( \frac{\ln p/\Lambda_{\rm QCD}}{\ln \mu/\Lambda_{\rm QCD}}\right)^{-\frac{3}{4}\frac{N_c}{2b}} \\
       \end{pmatrix}
        {\bf D}_{t}(p) .
\end{align}

This result for the ghost part is new, while the gluon part reproduces the recent result of Boucaud et. al. \cite{Boucaudetal01} 
in the MOM scheme (Note that their definition of $\gamma$ is different from ours by a factor 2 and the 
coefficient $\gamma_0$ differs by the signature). 
In order to transfer from our renormalization scheme to the MOM scheme, we have used the translation rule (\ref{rule}).
In the Landau gauge, therefore, we have confirmed that the ghost condensation does not affect the inverse gluon propagator as in the tree level, even if the leading logarithmic corrections are taken into account in the OPE.  In other words, the gluon condensation is decoupled from the ghost condensation within this approximation.

\subsection{Full propagators: momentum dependence}
\par
The vacuum polarization tensor of the gluon is decomposed into the transverse and the 
longitudinal parts:
\begin{equation}
  \Pi^{AB}_{\mu\nu}(p) =
    [\Pi^{\rm T}(p^2) P^{\rm T}_{\mu\nu}  +  \Pi^{\rm L}(p^2) P^{\rm L}_{\mu\nu} ] \delta^{AB} , 
\end{equation}
where $\Pi^{\rm T}$ and $\Pi^{\rm L}$ are functions of $p^2$ alone.
Once the vacuum polarization functions $\Pi^{\rm T}$ and $\Pi^{\rm L}$ of the gluon are obtained from the OPE,  the propagator is written as
\begin{align}
  (D)^{AB}_{\mu\nu}(p) 
 &= \delta^{AB} \left[ {1 \over -p^2+\Pi^{\rm T}(p^2)} P^{\rm T}_{\mu\nu} + {\lambda \over -p^2+\lambda 
\Pi^{\rm L}(p^2)}  P^{\rm L}_{\mu\nu} \right] 
\\
&= \delta^{AB} \left[ {Z_{\rm gl}(-p^2) \over -p^2} P^{\rm T}_{\mu\nu} + {\lambda \over -p^2+\lambda 
\Pi^{\rm L}(p^2)}  P^{\rm L}_{\mu\nu} \right] ,
\end{align}
where we have defined a function $Z_{\rm gl}(-p^2)$ by
\begin{equation}
  Z_{\rm gl}(-p^2) = Z_{\mbox{\tiny pert}}(-p^2) + Z_{\rm OPE}(-p^2) := {-p^2 \over -p^2+\Pi^{\rm T}(p^2)} .
\end{equation}
Note that $\Pi^{\rm L}(p^2) \equiv 0$ in the conventional Landau gauge.
\par
On the other hand, if the vacuum polarization function of the ghost 
$\Pi_{\rm gh}^{AB}(p^2)=\delta^{AB}\Pi_{\rm gh}(p^2)$ is calculated by OPE, the ghost propagator is 
obtained as
\begin{equation}
  G^{AB}(p) = [(G_0)^{-1}+i\Pi_{\rm gh}(p^2)]^{-1}_{AB} = {1 \over -ip^2 + i\Pi_{\rm gh}(p^2)} 
\delta^{AB} 
= (-i) {G_{\rm gh}(-p^2) \over -p^2} \delta^{AB} ,
\end{equation}
where we have introduced a function $G_{\rm gh}(-p^2)$ by
\begin{equation}
  G_{\rm gh}(-p^2) = G_{\mbox{\tiny pert}}(-p^2) + G_{\rm OPE}(-p^2) := {-p^2 \over -p^2 + \Pi_{\rm gh}(p^2)} .
\end{equation}
\par

The OPE contribution $\bm{\Pi}^{\rm OPE}$ to the vacuum polarization function in the inverse propagators (\ref{invgluonOPE}) and (\ref{invghostOPE}) is related to the Wilson coefficient ${\bf C}^{\rm 1PI}$ as 
\begin{equation}
  \bm{\Pi}^{\rm OPE} :=  
  \begin{pmatrix}
   \Pi_{\rm gl}^{\rm OPE} \\
   \Pi_{\rm gh}^{\rm OPE}
  \end{pmatrix}
 =  {\bf C}^{\rm 1PI} \begin{pmatrix}
    \frac12 A^2 \\
    i\bar C C
  \end{pmatrix} 
= 
   \begin{pmatrix}
     (iD_{\mbox{\tiny pert}})^{-2}  & 0 \\
     0 & (iG_{\mbox{\tiny pert}})^{-2} 
   \end{pmatrix} {\bf F} 
  \begin{pmatrix}
    \frac12 A^2 \\
    i\bar C C
  \end{pmatrix} .
\label{Pi}
\end{equation}
Substituting the result (\ref{recomb}) into (\ref{Pi}), we obtain a pair of vacuum polarization functions:
\begin{align}
  \bm{\Pi}^{\rm OPE}(p) = 
      \begin{pmatrix}
          ( T_2-\frac{13}{3}T_1) 
           (\frac{\ln p/\Lambda_{\rm QCD}}{\ln \mu/\Lambda_{\rm QCD}})^{\frac{35}{12}\frac{N_c}{2b}} \frac{1}{(iD_{\mbox{\tiny pert}})^{2}}
       &  ( T_2-\frac{26}{3}T_1) 
           (\frac{\ln p/\Lambda_{\rm QCD}}{\ln \mu/\Lambda_{\rm QCD}})^{\frac{61}{12}\frac{N_c}{2b}} \frac{1}{(iD_{\mbox{\tiny pert}})^{2}}
           \\
           -\frac{13}{3}T_3  
           (\frac{\ln p/\Lambda_{\rm QCD}}{\ln \mu/\Lambda_{\rm QCD}})^{\frac{13}{6}\frac{N_c}{2b}} \frac{1}{(iG_{\mbox{\tiny pert}})^{2}}
       &   -\frac{26}{3}T_3 
           (\frac{\ln p/\Lambda_{\rm QCD}}{\ln \mu/\Lambda_{\rm QCD}})^{\frac{13}{3}\frac{N_c}{2b}} \frac{1}{(iG_{\mbox{\tiny pert}})^{2}}
      \end{pmatrix}
  \begin{pmatrix}
    {\cal Q}_1 \\
    {\cal Q}_2 
  \end{pmatrix} .
\end{align}
It turns out that the vacuum polarization functions just obtained reduce to the tree results, i.e., (\ref{gluonVP}) and (\ref{ghostVP}), at $\kappa=1$ (or $p=\mu$).
Therefore, the ghost condensation $\langle i\bar{C}C\rangle$  does contribute to  the  gluon and ghost vacuum polarization functions in the leading logarithmic corrections of the OPE.
\par
Thus the OPE contribution to the gluon and ghost vacuum polarization functions are obtained:
\begin{align}
   \Pi_{T}{}^{\rm OPE}(p^2) =& \frac{2\pi^2}{b}\frac{ N_c (1+\lambda)}{(N_c^2-1)}
        \Bigg\{ 
            \frac{ {(\ln \frac{p}{\Lambda_{\rm QCD}})}^{\frac{35}{12}\frac{N_c}{2b} -1} }
                  { {(\ln \frac{\mu}{\Lambda_{\rm QCD}})}^{\frac{35}{12}\frac{N_c}{2b}} }
         \left(  \langle \frac12 A^2\rangle \nonumber 
   +   \frac{26}{3}  \langle i \bar CC \rangle \right)
 \\
 & -  2 \frac{ {(\ln \frac{p}{\Lambda_{\rm QCD}})}^{\frac{61}{12}\frac{N_c}{2b} -1} }
                  { {(\ln \frac{\mu}{\Lambda_{\rm QCD}})}^{\frac{61}{12}\frac{N_c}{2b}} }
          \left(   \langle \frac12 A^2\rangle  
    +      \frac{13}{3} \langle i \bar CC \rangle   \right)       
        \Bigg\} \left(\frac{D_0(p)}{D_{\mbox{\tiny pert}}(p)}\right)^2 ,
        \label{gluonPi}
\end{align}
\begin{align}
    \Pi_{\rm gh}^{\rm OPE}(p^2) =& \frac{2\pi^2}{b}\frac{N_c}{(N_c^2-1)}
        \Bigg\{ 
            -\frac{ {(\ln \frac{p}{\Lambda_{\rm QCD}})}^{\frac{13}{6}\frac{N_c}{2b} -1} }
                  { {(\ln \frac{\mu}{\Lambda_{\rm QCD}})}^{\frac{13}{6}\frac{N_c}{2b}} }
         \left(  \langle \frac12 A^2\rangle \nonumber 
   +   \frac{26}{3}  \langle i \bar CC \rangle \right)
 \\
 & +  2 \frac{ {(\ln \frac{p}{\Lambda_{\rm QCD}})}^{\frac{13}{3}\frac{N_c}{2b} -1} }
                  { {(\ln \frac{\mu}{\Lambda_{\rm QCD}})}^{\frac{13}{3}\frac{N_c}{2b}} }
          \left(  \langle \frac12 A^2\rangle  
    +      \frac{13}{3} \langle i \bar CC \rangle   \right)       
        \Bigg\} \left(\frac{G_0(p)}{G_{\mbox{\tiny pert}}(p)}\right)^2 .
         \label{ghostPi}
\end{align}

\par
The effective gluon mass is obtained from the pole of $Z_{\rm gl}(-p^2)$, i.e., a solution of the 
equation $p^2=\Pi_{\rm T}(p^2)$, while the effective ghost mass is obtained from the pole of $G_{\rm gh}
(-p^2)$, i.e., a solution of the equation $p^2=-i\Pi_{\rm gh}(p^2)$.  
In view of this, the solutions (\ref{gluonPi}) and (\ref{ghostPi}) 
would give an improvement of the tree-level result, 
(\ref{gluonmass}) and (\ref{ghostmass}).  
However, a BRST non-invariant combination ${\cal Q}_2$ of composite operators appears together with the BRST invariant combination ${\cal Q}_1$ discussed in the previous section.  Therefore, these results indicate that we need more endeavor in order to reach the BRST invariant pole position in the IR region.
\par
In the Landau gauge, especially, we have
\begin{align}
  Z_{\rm gl}(-p^2) &= -p^2 D_{\mbox{\tiny pert}}(p)  -p^{-2} \left(
          \frac{\pi^2}{b}\frac{N_c}{(N_c^2-1)} 
          \frac{ {(\ln \frac{p}{\Lambda_{\rm QCD}})}^{\frac{3}{4}\frac{N_c}{2b} -1} }
               { {(\ln \frac{\mu}{\Lambda_{\rm QCD}})}^{\frac{3}{4}\frac{N_c}{2b}} }
          \langle A^2\rangle
                 \right),
\\
  G_{\rm gh}(-p^2) &= -ip^2 G_{\mbox{\tiny pert}}(p)  +p^{-2} \left(
          \frac{\pi^2}{b}\frac{N_c}{(N_c^2-1)} 
          \frac{ {(\ln \frac{p}{\Lambda_{\rm QCD}})}^{\frac{13}{6}\frac{N_c}{2b} -1} }
               { {(\ln \frac{\mu}{\Lambda_{\rm QCD}})}^{\frac{13}{6}\frac{N_c}{2b}} }
          \langle A^2\rangle
                 \right).
\end{align}
After the Wick rotation to the Euclidean region 
$p^2 \rightarrow -p_E^2$, we find that the function $Z_{\rm gl}(p_E^2)$ is monotonically 
increasing in $p_E^2$ if 
$\langle A_E^2 \rangle := - \langle A^2 \rangle > 0$, as in the case of constant 
$\Pi^{\rm T}(-p_E^2)=M^2>0$.
On the other hand, if 
$\langle A_E^2 \rangle := - \langle A^2 \rangle < 0$, $Z_{\rm gl}(p_E^2)$ has a Landau pole in the 
IR region and is monotonically decreasing in $p_E^2$ in the UV region.
In the conventional Landau gauge, these results can be compared with those of the 
Schwinger-Dyson equation (see e.g., \cite{AS01}) and the numerical simulation on a lattice (see e.g., 
\cite{Boucaudetal01,BBLSW98,SS96,LRG01}).  According to these results, $Z_{\rm gl}(p_E^2)$ is 
enhanced at intermediate momenta and has a peak at about 1 GeV.  
It was argued \cite{LRG01} that the enhancement of the gluonic form factor at IR region is 
related to quark confinement.  
However, this region is beyond the reach of our study in this paper.  
Incidentally, the data in the gauge other than the Landau gauge is not yet available.

\section{Conclusion and discussion}

In this paper we have discussed the multiplicative renormalizability of the composite operator 
$\mathcal{O}$ in QED and Yang-Mills theory.
This research is motivated by clarifying the mechanism of mass generation 
and a possible connection to quark confinement.
\par
In QED, we have shown that the composite operator is trivially renormalizable and that the renormalized 
composite operator is BRST and anti-BRST invariant for an arbitrary value of the gauge fixing 
parameter.  There is no subtlety related to the renormalization of the composite operator.
\par
In the Yang-Mills theory, we have adopted the most general Lorentz gauge with two gauge-fixing 
parameters $\xi, \lambda$ which was derived by the Baulieu and Thierry-Mieg \cite{BT82}. 
It was known \cite{Kondo01} that the bare composite operator $\mathcal{O}$ of mass dimension 2 
is invariant under the {\it bare} BRST and anti-BRST transformation for the choice of gauge 
parameters $\lambda=0$ or $\xi={1 \over 2}$ and that it is also invariant under the gauge 
transformation in the Landau gauge $\lambda=0$. 
In this paper the composite operator has been renormalized by taking into account the operator mixing 
carefully.  Here the matrix of renormalization factors has  been explicitly calculated. 
Consequently, we have found that the BRST and anti-BRST invariance of the renormalized 
composite operator $\mathcal{O}^{\rm R}$ holds if the renormalized parameters take the same 
value, $\lambda_{\rm R}=0$ or $\xi_{\rm R}={1 \over 2}$, as the bare one.  
Moreover, we have obtained the RG flow in the $(\xi,\lambda)$ plane to one-loop order.
In the RG flow diagram, the RG flow runs only on the line $\xi_{\rm R}={1 \over 2}$ if the 
initial position of $\xi$ is located somewhere on the line. The line $\lambda_{\rm R}=0$ is a 
line of fixed points.  Therefore, if the system is located on a point in the line $\lambda_{\rm R}=0$ initially, it can not move from 
the initial position.  This fact guarantees the BRST invariance of the renormalized composite operator $\mathcal{O}^{\rm R}$.  
\par
  We have also examined in this paper how the conventional calculations are modified in the presence of the vacuum condensate of mass dimension 2.  
  By performing the OPE of the gluon and ghost propagators, we have shown that the effective masses of gluon and ghost are generated due to the non-vanishing vacuum condensate.  Although this phenomenon was already suggested based on the tree level calculation, 
  we have taken into account the leading logarithmic corrections in consistent with the RG flow by making use of the RG equation.
  We have found that  the effective masses are provided from the ghost condensation $\langle i \bar{\mathscr{C}} \cdot \mathscr{C} \rangle$ as well as the gluon condensation $\langle {1 \over 2}\mathscr{A}_\mu \cdot \mathscr{A}^\mu \rangle$ (except for the Landau gauge $\lambda=0$). This result should be compared with the tree level result where the effective mass has the contribution from the gluon condensate alone.
\par
The next step is to show that the non-vanishing vacuum condensates 
$\langle \mathcal{O} \rangle \not=0$ is actually realized in the QCD vacuum. 
An attempt in this direction has already been performed 
in \cite{KS00} by calculating the effective potential for the ghost condensation 
$\langle i \bar{C}C \rangle$ in the $SU(2)$ and $SU(3)$ Yang-Mills theories in the MA gauge. 
Quite recently, Verschelde et al.\cite{VKAV01} have carefully obtained the multiplicatively 
renormalizable effective potential for the gluon condensate
$\langle {1 \over 2}\mathscr{A}_\mu \mathscr{A}^\mu \rangle$ in the Landau gauge up to 
two-loop order in the $SU(N)$ Yang-Mills theory.  Both results support that the non-zero vacuum 
condensate of mass dimension 2 is energetically favoured in the Yang-Mills theory.
In these approaches, an auxiliary field $\rho(x)$ corresponding to the composite operator has 
been introduced  to obtain the effective potential $V(\sigma)$ of a constant $\sigma=\rho(x)$.  
However, this treatment has a number of subtle points which have not been discussed in these 
papers.    
This issue will be discussed in a subsequent paper \cite{KIMS01} in detail.  
\par
In massless QED, photon pairing \cite{Fukuda89,IF91} can occur in the strong coupling phase 
\cite{Miransky85,BLL86,KMY89} where the chiral symmetry is spontaneously broken.  Therefore, 
it will be possible to discuss the interplay between quark confinement and chiral symmetry 
breaking on equal footing in a unified treatment.  The extension of this viewpoint into the 
non-Abelian case, i.e., gluon pairing \cite{KSY91} is also an interesting subject to be 
tackled in the future work.  
\par
Finally, we point out that the operator $\mathcal{O}$ is essentially a mass term for the gluon 
and the ghost fields.  
Although a naive introduction of a mass term for the gluon alone breaks the BRST symmetry,
our result indicates that there is a BRST invariant combination of mass terms:
\begin{equation}
 \mathscr{L}_m :=\text{tr} \left[ 
 {1 \over 2} m^2 \mathscr{A}_\mu(x) \cdot \mathscr{A}_\mu(x) + m^2 \alpha i 
\bar{\mathscr{C}}(x) \cdot \mathscr{C}(x) 
\right] .
\label{mass}
\end{equation}
This mass term is very similar to that obtained after the spontaneous breakdown caused by the 
non-vanishing vacuum expectation value of the Higgs scalar field.  In our case, the mass 
should be of dynamical origin.  It is possible to give a proof of the multiplicative 
renormalizability of the Yang-Mills theory with a mass term preserving the BRST symmetry to 
all orders of perturbation theory.
However, it is known \cite{DV70,Ojima82} that the introduction of the mass term (\ref{mass}) 
breaks the nilpotency of the off-shell BRST transformation as well as the on-shell one.  
Consequently, the unitarity of the theory turns out to be spoiled.  In this sense, the mass generation 
should occur in the dynamical way, i.e., 
$\langle \mathcal{O} \rangle \not=0$ in the limit $m \rightarrow 0$.
This viewpoint will be discussed in a subsequent paper.
\section*{Acknowledgments}
One of the authors (K.-I. K.) would like to thank V.P. Gusynin for informing him of the old works 
\cite{Fukuda78,GM82} on the instability of the QCD vacuum caused by gluon pairing.  
This work is supported by Sumitomo Foundations and  in part by 
a Grant-in-Aid for Scientific Research from the Ministry of
Education, Science and Culture:(B) No.13135203.

\appendix
\section{OPE calculations}


In order to give the OPE correction for the gluon propagator, we need to calculate the 
following diagrams.
\begin{align}
  \begin{array}{c}
   \includegraphics[width=1.4cm]{OPEaa-a_a1.eps}
  \end{array}
   =\,& \frac{1}{(N_c^2-1)D} g^{\rho\rho'} \delta^{CC'} g f^{ACD} 
[g_{\mu\sigma}(-2p)_\rho+g_{\sigma\rho}p_\mu+g_{\rho\mu}p_\sigma] \nonumber\\
  & \times 
       \frac{-i}{p^2}(P^{\rm T}_{\sigma\sigma'}+\lambda P^{\rm L}_{\sigma\sigma'})\delta^{DD^\prime}
   g f^{BD'C'} [g_{\nu\sigma'}(-2p)_{\rho'}+g_{\sigma'\rho'}p_\nu+g_{\rho'\nu}p_{\sigma'}] 
\nonumber \\
  =\,& \frac{1}{(N_c^2-1)D} i g^2 (-N_c) \frac{1}{p^2} [(4+\lambda)p^2 P^{\rm T}_{\mu\nu} +(D-1)p^2 
P^{\rm L}_{\mu\nu}] \delta^{AB} .
\end{align}

\begin{align}
  \begin{array}{c}
   \includegraphics[width=1.0cm]{OPEaa-a_a3.eps}
  \end{array}
  =\,& \frac{1}{(N_c^2-1)D} i g^2 g^{\gamma\delta}\delta^{CD} \big[
   f^{EAB}f^{ECD}(g_{\mu\gamma}g_{\nu\delta}-g_{\mu\delta}g_{\nu\gamma})\nonumber\\
& +f^{EAC}f^{EBD}(g_{\mu\nu}g_{\gamma\delta}-g_{\mu\delta}g_{\gamma\nu})
  +f^{EAD}f^{EBC}(g_{\mu\nu}g_{\gamma\delta}-g_{\mu\gamma}g_{\delta\nu})\big]\nonumber\\
             =\,& ig^2 \frac{2}{(N_c^2-1)D} N_c [g_{\mu\nu}(D-1)]\delta^{AB} .
\end{align}

\begin{align}
  \begin{array}{c}
   \includegraphics[width=1.4cm]{OPEcc-a_a1.eps}
  \end{array}
   =\,&  \frac{1}{(N_c^2-1)} (p_\mu+\xi(-p)_\mu) g f^{ADC} \frac{-1}{p^2}\delta^{CC'} (0+\xi 
p_{\nu}) g f^{BC'D'}\delta^{DD'} \nonumber\\
  =\,&  \frac{N_c g^2}{(N_c^2-1)} \frac{1}{p^2} \xi(1-\xi) p_\mu p_{\nu} \delta^{AB} .
\end{align}


For the correction of the ghost propagator, we need the calculation of the following diagrams.
\begin{align}
  \begin{array}{c}
  \includegraphics[width=1.8cm]{OPEaa-c_c1.eps}
  \end{array}
     =\,& \frac{\delta^{CC'} g_{\mu\nu}}{(N_c^2-1)D}i g f^{CAD} (p_\mu) 
\frac{-1}{p^2}\delta^{DD'} i g f^{C'D'B}(p_\nu) \nonumber \\
   =\,& -\frac{N_c g^2}{(N_c^2-1)D}\delta^{AB}\ .
\end{align}

\begin{align}
  \begin{array}{c}
  \includegraphics[width=1.4cm]{OPEcc-c_c.eps}
  \end{array}
   =\,& - \frac{\delta^{CD}}{(N_c^2-1)}  (- i g^2) \lambda\xi(1-\xi) 
(f^{EAB}f^{EDC}-f^{EAC}f^{EDB}) \nonumber\\
   =\,& i \frac{N_c g^2}{(N_c^2-1)} \lambda\xi(1-\xi) \delta^{AB}.
  \label{cc2cc}
\end{align}

\begin{align}
  \begin{array}{c}
  \includegraphics[width=1.8cm]{OPEcc-c_c2.eps}
  \end{array}
 =\,&  \frac{\delta^{DD'}}{(N_c^2-1)} i g f^{DAC} (- p_\mu)(1-\xi)
       i \frac{-1}{p^2}(P_{\rm T} + \lambda P_{\rm L})^{\mu\nu}\delta^{CC'} i g f^{D'C'B}(\xi(-p_\nu)) 
\nonumber \\
  =\,& - i \frac{N_c g^2}{(N_c^2-1)} \lambda \xi (1-\xi) \delta^{AB}\ .
\end{align}


\end{document}